\documentclass[journal]{IEEEtran}

\IEEEoverridecommandlockouts
\usepackage{amsmath}

\usepackage{amsmath}
\usepackage{booktabs}
\usepackage{cite}
\usepackage{calc}
\usepackage{amsmath,amssymb,amsfonts}
\usepackage{mathrsfs}
\usepackage{graphicx}
\usepackage{textcomp}
\usepackage{xcolor}
\usepackage{mathtools}
\usepackage{amsmath} 
\usepackage{amssymb}
\usepackage{amsfonts}
\usepackage{verbatim} 
\usepackage{bm,color,soul}
\usepackage{algorithm}
\usepackage{algorithm}
\usepackage{algorithmicx}
\usepackage{algpseudocode}
\usepackage{array}
\usepackage[caption=false,font={footnotesize,rm},labelfont=rm,textfont=rm]{subfig}
\usepackage{url}
\usepackage{balance}
\usepackage{hyperref}
\usepackage{mathtools}
\usepackage{amsmath} 
\usepackage{amssymb}
\usepackage{amsfonts}
\usepackage{verbatim} 
\usepackage{bm,color,soul}
\usepackage{cite}
\usepackage{stfloats} 
\usepackage{subeqnarray}
\usepackage{cases}
\usepackage{amsthm}
\usepackage{makecell} 
\usepackage{enumerate}
\usepackage{multirow}
\usepackage{enumitem}

\newcommand{\rnum}[1]{\uppercase\expandafter{\romannumeral #1\relax}}
\providecommand{\lemmaname}{Lemma}
\providecommand{\propositionname}{Proposition}
\providecommand{\remarkname}{Remark}

\usepackage{bm}
\ifCLASSINFOpdf
\else
\fi

\begin{document}
%
\title{6D Channel Knowledge Map Construction via Bidirectional Wireless Gaussian Splatting}
%
%
%

\author{Juncong~Zhou,
        Chao~Hu,
        Guanlin~Wu,
        Zixiang~Ren,
        Han~Hu,
        Juyong~Zhang,
        Rui~Zhang,~\IEEEmembership{Fellow,~IEEE,}
        and~Jie~Xu,~\IEEEmembership{Fellow,~IEEE}
\thanks{J. Zhou, C. Hu, G. Wu, Z. Ren, and J. Xu  are with the
 School of Science and Engineering (SSE), the Shenzhen Future Network of Intelligence Institute (FNii-Shenzhen), and the Guangdong
 Provincial Key Laboratory of Future Networks of Intelligence, The
 Chinese University of Hong Kong, Shenzhen, Guangdong 518172,
 China (e-mail: juncongzhou@link.cuhk.edu.cn; chaohu@link.cuhk.edu.cn; guanlinwu1@link.cuhk.edu.cn; rzx66@mail.ustc.edu.cn; xujie@cuhk.edu.cn). J. Xu is the corresponding author.}
\thanks{H. Hu is with the School of Information and Electronics, Beijing Institute
 of Technology, Beijing 100081, China (e-mail: hhu@bit.edu.cn).}
\thanks{J. Zhang is with the School of Mathematical Sciences,
 University of Science and Technology of China, Hefei 230027,
 China. (e-mail: juyong@ustc.edu.cn).}
\thanks{ R. Zhang is with  the Department
 of Electrical and Computer Engineering, National University of Singapore,
 Singapore 117583 (e-mail: elezhang@nus.edu.sg).}}

\maketitle

\begin{abstract}
This paper investigates the construction of channel knowledge map (CKM) from sparse channel measurements. Different from conventional two-/three-dimensional (2D/3D) CKM approaches assuming fixed base station configurations, we present a six-dimensional (6D) CKM framework named bidirectional wireless Gaussian splatting (BiWGS), which is capable of modeling wireless channels across dynamic transmitter (Tx) and receiver (Rx) positions in 3D space. BiWGS uses Gaussian ellipsoids to represent virtual scatterer clusters and environmental obstacles in the wireless environment. By properly learning the bidirectional scattering patterns and complex attenuation profiles based on channel measurements, these ellipsoids inherently capture the electromagnetic transmission characteristics of wireless environments, thereby accurately modeling signal transmission under varying transceiver configurations. Experiment results show that BiWGS significantly outperforms classic multi-layer perception (MLP) for the construction of 6D channel power gain map with varying Tx-Rx positions, and achieves spatial spectrum prediction accuracy comparable to the state-of-the-art wireless radiation field Gaussian splatting (WRF-GS) for 3D CKM construction. This validates the capability of the proposed BiWGS in accomplishing dimensional expansion of 6D CKM construction, without compromising fidelity.

\end{abstract}


%
\IEEEpeerreviewmaketitle

\section{Introduction}
%
%
%
%
Acquisition of accurate channel state information (CSI) is becoming increasingly important for resource allocation and beamforming optimization in wireless communication systems. This holds particular significance for future sixth-generation (6G) networks with ultra-dense base station (BS) deployment, extremely large-scale antenna array (ELAA), and ultra-high bandwidth \cite{ren2024sensing,fang2017joint}. Conventionally, real-time CSI acquisition is achieved through pilot-based channel estimation and limited feedback, which, however, may cause prohibitive signaling overhead\cite{giordani2018tutorial}. Recently, channel knowledge map (CKM) has emerged as a promising solution to tackle this challenge \cite{zeng2021toward}. 
CKM provides {\it a priori} channel knowledge (e.g., channel power gain\cite{sun2025channel}, beam index\cite{wu2021environment}, channel angle\cite{wu2023environment}, and CSI), enabling environment-aware communication while reducing and even eliminating the requirements of real-time channel measurements. In particular, CKM can be represented in the forms of a spatial database, images, or neural network models, which constitutes a mapping relationship from wireless environments and spatial positions of transceivers to channel knowledge. CKM is envisioned to enable a wide range of environment-aware applications, including predictive communication, resource allocation, beam tracking, unmanned aerial vehicle (UAV) placement, as well as sensing and localization\cite{zeng2024tutorial}.

CKMs can be categorized as base station (BS)-to-any (B2X) and any-to-any (X2X) CKMs, depending on the input dimension of transceiver positions\cite{zeng2021toward}. In particular, the B2X CKM utilizes the two-dimensional (2D) or three-dimensional (3D) position of mobile user as input, providing channel knowledge at that specific position relative to a fixed-position BS, thereby supporting BS-centric communications. By contrast, the X2X CKM exploits both the transmitter (Tx) and receiver (Rx) positions as input, providing the channel knowledge at varying Tx and Rx positions, which makes it suitable for X2X and device-to-device (D2D) communications. Depending on whether 2D or 3D space is considered, the X2X CKMs can be constructed in 4D (with 2D Tx/Rx positions) or 6D (with 3D Tx/Rx positions) formats, respectively. In particular, the 6D X2X CKM provides the most comprehensive wireless environment/channel information, which is crucial for solving complex tasks such as environment-aware BS deployment \cite{xia2023multiple} and 3D trajectory planning for low-altitude UAVs\cite{zhang2020radio}.

The construction of CKM is essential for its practical implementation. In the literature, various methods for CKM construction have been proposed. The representative ones include ray tracing\cite{suga2020ray,hoydis2023sionna}, interpolation\cite{zhang2022k,sun2022propagation,hu2020spatiotemporal}, deep learning \cite{saito2019two,levie2021radiounet,wang2024deep}, and wireless radiation field (WRF)\cite{zhao2023nerf2,lu2024newrf,wen2025wrf}. First, the ray tracing approach enables high-accuracy channel reconstruction by leveraging physical environment information. It models electromagnetic (EM) waves as particles and simulates their interactions with the environment through reflection, scattering, and diffraction\cite{suga2020ray,hoydis2023sionna}. Ray tracing achieves high accuracy when precise environmental geometry and material properties are known, but its practical application is hindered by the extreme computational complexity and the difficulty in acquiring complete environment knowledge {\it a priori}. Next, the interpolation approaches aim to reconstruct channel knowledge at unmeasured positions by using limited channel measurements at reference positions via techniques like K-nearest neighbors (KNN) \cite{zhang2022k}, matrix completion \cite{sun2022propagation}, and Kriging interpolation \cite{hu2020spatiotemporal}, in which their spatial correlations and relative distances are utilized. However, the interpolation-based methods rely on the assumption of fixed Tx or Rx positions, and are highly dependent on the stationarity of the environment without accounting for the environmental geometric structure.  Therefore, these methods are less effective in dynamic or complex scenarios, and are only suitable for constructing 2D/3D B2X CKMs but not applicable for 4D/6D X2X CKMs. Furthermore, deep learning approaches have also been used as an efficient approach for CKM construction due to their ability to learn complex non-linear mappings from limited measurement data. A typical example is multi-layer perception (MLP) \cite{saito2019two}, which directly learns the channel features from the Tx-Rx position pair inputs to predict the channel power gain. However, the MLP method lacks the capacity to incorporate geometric information of the environment, resulting in significant performance degradation in complex environments. Beyond MLPs, deep learning-based image processing techniques have also been employed (see, e.g., \cite{levie2021radiounet,wang2024deep}). The authors in \cite{levie2021radiounet} proposed a convolutional neural network (CNN)-based method called RadioUnet, which uses the Tx positions and city maps as inputs to estimate channel gains at arbitrary Rx positions in the city, thereby constructing a channel gain map. Furthermore, the authors in \cite{wang2024deep} introduced a super-resolution (SR)-based CKM construction method, in which the SR residual network (SRResNet) is employed to recover a high-quality CKM image from sparse, low-resolution observation data. However, the image processing-based approaches are proposed for 2D B2X and 4D X2X CKMs, not applicable for 6D X2X CKMs of our interest. 

Recently, the WRF approaches have emerged as a new approach for CKM construction. The emergence of WRF approaches is inspired by the recent advances of radiance field rendering techniques (especially the Neural Radiance Fields (NeRF)\cite{mildenhall2020nerf} and 3D Gaussian Splatting (3DGS) \cite{kerbl20233d}) for 3D scene reconstruction in the computer graphics field. While NeRF implicitly represents the radiance field as an MLP that is trained from a few images to synthesize novel-view images, 3DGS explicitly represents the radiance field as Gaussian ellipsoids colored via spherical harmonics (SH). In practice, NeRF is relatively time-consuming in both training and inference \cite{muller2022instant}, while 3DGS can achieve high-resolution rendering at faster speeds through tile-based rendering. With the great success of NeRF and 3DGS in computer graphics and motivated by the fact that both light and wireless signals are EM waves, the NeRF and 3DGS techniques have recently been employed as effective WRF approaches for CKM construction. For instance, a NeRF-based WRF method named neural radio-frequency radiance fields (NeRF$^2$) \cite{zhao2023nerf2} was proposed for spatial spectrum prediction for a single-input multiple-output (SIMO) wireless communication system with a signal-antenna mobile user as Tx and a multi-antenna fixed-position BS as Rx. NeRF$^2$ employs the 3D Tx position, ray direction with respect to Rx, and voxel position (position of sampled points along the transmission ray) as the inputs of MLP, which fits the amplitude attenuation and phase shifts of wireless signals in specific receive directions, thereby constructing a 3D B2X CKM. To reduce the computational complexity of NeRF$^2$, a follow-up work \cite{lu2024newrf} developed Neural Wireless Radiance Fields (NeWRF), which estimates the angle of arrival (AOA) of received signals via spatial signal classification algorithms such as multiple signal classification (MUSIC), thus minimizing the number of generated rays for tracing and accordingly reducing the computation complexity for training and inference. On the other hand, building upon 3DGS, wireless radiation field Gaussian splatting (WRF-GS) was proposed in \cite{wen2025wrf} to use 3D Gaussian ellipsoids to model the WRF, which capture the electromagnetic transmission characteristics from received signals, thus improving the accuracy of 3D CKM construction at enhanced computational efficiency. Despite the progress in using NeRF and 3DGS for CKM construction, these prior works focused on the scenario when the Rx position is static (e.g., when Rx is a BS), thus making them suitable for 3D B2X CKMs, but not for 6D X2X CKMs. 

Different from prior WRF works focusing on 3D B2X CKM construction, in this paper we consider the construction of 6D X2X CKM for a SIMO system with varying 3D Tx-Rx positions. In particular, we propose a new framework named \emph{Bidirectional} Wireless Gaussian Splatting (BiWGS), which uses Gaussian ellipsoids to represent virtual scatterer clusters as well as environmental obstacles. BiWGS is inspired by the Bidirectional Gaussian splatting (BiGS) algorithm \cite{zhenyuan2025bigs} in the computer graphics field, which is a 3DGS-based design for 3D scene reconstruction under varying illumination. Different from conventional 3DGS, BiGS uses bidirectional spherical harmonics (BSH) function to model the optical bidirectional scattering pattern of Gaussian ellipsoids based on the AOA and angle of departure (AOD), capturing the dynamic illumination. Motivated by this, the proposed BiWGS incorporates the idea of BiGS into CKM construction, in which the parameters of Gaussian ellipsoids and BSHs are properly learned to capture the \emph{bidirectional} scattering patterns and complex attenuation profiles under different Tx-Rx position pairs, thereby enabling the efficient construction of 6D CKM.

The main results of this paper are listed as follows.

\begin{itemize}
    \item First, we adopt a scatterer-cluster-based channel model to facilitate the BiWGS design.  This model represents the wireless channel between any Tx-Rx pair as a multi-path channel comprising multiple scattering paths and a potential line-of-sight (LOS) path, each subject to complex attenuations caused by environmental obstacles along the paths. Building on this channel modeling, we propose the BiWGS framework, which leverages Gaussian ellipsoids to serve as both virtual scatterer clusters and environmental obstacles, effectively representing the wireless environment. Notably, each Gaussian ellipsoid is characterized by a bidirectional scattering profile, enabling the construction of 6D X2X CKMs.
 
    \item Then, we adopt the Gaussian ellipsoid representation to model the attenuation of SIMO channel paths. Initially, the BSH function is used for each Gaussian ellipsoid to represent the bidirectional complex scattering coefficient of its associated virtual scatterer clusters. Subsequently, for each channel path, wireless splatting is employed to compute the complex attenuation induced by the obstruction of each Gaussian ellipsoid, and these individual attenuations are aggregated via wireless rendering to reconstruct the attenuation over the path. By combining the channel paths from all possible AOAs at Rx, we obtain the SIMO channel vector between any Tx-Rx position pair.
    
    \item In the training process, we use the weighted sum of the spatial spectrum prediction loss and the channel power gain loss as the loss function of BiWGS, in which the spatial spectrum represents the angular power distribution of the wireless channel. In addition, an adaptive density control strategy is applied during the backward propagation for training, dynamically adjusting the number of Gaussian ellipsoids and their sizes based on the gradient magnitudes.
    
    \item Finally, we provide experiment results to validate the effectiveness of our method in CKM construction. The proposed BiWGS demonstrates significant performance gains over the benchmark MLP scheme in constructing 6D channel power gain maps across varying Tx-Rx pairs. Furthermore, in 3D B2X CKM construction, BiWGS achieves spatial spectrum prediction accuracy on a par with the state-of-the-art (SOTA) benchmark WRF-GS. These results demonstrate that BiWGS effectively achieves dimensional expansion from 3D B2X CKM to 6D X2X CKM, while maintaining high fidelity.
    
\end{itemize}

The remainder of this paper is organized as follows. Section \ref{Review of BiGS} reviews the basics of BiGS for 3D scene reconstruction. Section \ref{framework} presents the BiWGS framework for 6D X2X CKM construction. Section \ref{results} presents experiment results. Finally, Section \ref{conclusion} concludes the paper.

\textit{Notations}: Vectors and matrices are denoted by boldface lowercase and upper-case letters, respectively. For any vector or matrix, $(\cdot)^T$ and $(\cdot)^H$ refer to its transpose and conjugate transpose, respectively. $\boldsymbol{A}\otimes\boldsymbol{B}$ represents the Kronecker product of matrices $\boldsymbol{A}$ and $\boldsymbol{B}$. $\|\cdot \|_2 $ represents the Euclidean norm. $\mathbb{R}^{m\times n}$ and $\mathbb{C}^{m\times n}$ represent the spaces of real and complex matrices with dimension $m\times n$, respectively. The imaginary unit is denoted as $j = \sqrt{-1}$. $\lvert \ \cdot  \ \rvert$ denotes the amplitude of a complex number.

\section{Review of BiGS for 3D Scene Reconstruction with Dynamic Illumination }
\label{Review of BiGS}
Before we proceed to present our proposed BiWGS method, this section provides a brief review of BiGS for 3D scene reconstruction with varying illumination in the computer graphics field. 3DGS has demonstrated considerable success in reconstructing 3D scenes through collections of 3D Gaussian ellipsoids under static illumination. BiGS is an extension of the conventional 3DGS method. It reconstructs 3D scenes from images at limited input views and synthesizes images from novel views under varying illumination by learning the light-dependent color pattern of Gaussian ellipsoids.

\subsection{Bidirectional Gaussian Ellipsoid Representation}
First, we introduce the basic principle of BiGS scene representation. The modeling of Gaussian ellipsoids in BiGS follows the 3DGS methodology: It represents scenes via a set of anisotropic Gaussian ellipsoids, which are determined by 3D Gaussian distributions. These ellipsoids characterize spatially varying radiance fields in the environment, thus capturing both the geometric structure and visual appearance of a scene without the need for normal estimation\cite{kerbl20233d}. Moreover, BiGS incorporates the optical scattering function for each Gaussian ellipsoid to characterize its light-dependent color patterns under dynamic illumination, enabling scene representation under dynamic lighting conditions, a capability notably absent in the original 3DGS methodology. During the rendering process, these ellipsoids are projected onto the image plane via splatting for an arbitrary view. The final synthesized image is subsequently generated by employing the optical rendering equation, which aggregates the radiance contributions of all splatted 2D Gaussian ellipsoids per pixel through $\alpha$-blending.

Next, we explain the Gaussian ellipsoid representation for 3D scene in detail. For every Gaussian ellipsoid, its corresponding 3D Gaussian distribution is expressed as
\begin{equation}
G(\boldsymbol{x})=\exp\bigg\{-\frac{1}{2}(\boldsymbol{x}-\boldsymbol{\mu})^T\boldsymbol{\Sigma}^{-1}(\boldsymbol{x}-\boldsymbol{\mu})\bigg\},
\label{Gaussian_distribution}
\end{equation}
where $\boldsymbol{\mu}\in\mathbb{R}^{3}$ denotes the mean vector of ellipsoid, $\boldsymbol{\Sigma}\in\mathbb{R}^{3\times3}$ denotes the covariance matrix of ellipsoid governing its spatial extent and orientation, $\boldsymbol{x}\in\mathbb{R}^{3} $ denotes the 3D position, and the normalization constant of the Gaussian distribution is omitted. Moreover, the covariance matrix $\boldsymbol{\Sigma}$ in (\ref{Gaussian_distribution}) is determined by scaling matrix $\boldsymbol{S}\in\mathbb{R}^{3\times3}$ and rotation matrix $\boldsymbol{R}\in\mathbb{R}^{3\times3}$, i.e.,
\begin{equation}
    \boldsymbol{\Sigma}=\boldsymbol{R}\boldsymbol{S}\boldsymbol{S}^T\boldsymbol{R}^T.
    \label{covariance}
\end{equation}

BiGS models the light-dependent color pattern of each Gaussian ellipsoid through three key components. The first component is incident radiance $\boldsymbol{l}(\theta^{\prime},\phi^{\prime})\!\! =$ $\!\![l_\mathrm{R}(\theta^{\prime},\phi^{\prime}),l_\mathrm{G}(\theta^{\prime},\phi^{\prime}),l_\mathrm{B}(\theta^{\prime},\phi^{\prime})]^T$, which quantifies the amount of radiance ellipsoid received from the light source at the incident direction $(\theta^{\prime},\phi^{\prime})$, with $\theta^{\prime}$ and $\phi^{\prime}$ denoting the incident elevation and azimuth angles, respectively. The second component is optical scattering function $\boldsymbol{f}(\theta,\phi,\theta^{\prime},\phi^{\prime})=[f_\mathrm{R}(\theta,\phi,\theta^{\prime},\phi^{\prime}),f_\mathrm{G}(\theta,\phi,\theta^{\prime},\phi^{\prime}),$ $f_\mathrm{B}(\theta,\phi,\theta^{\prime},\phi^{\prime})]^T$, which characterizes the bidirectional scattering properties of ellipsoid. Here, $\theta$ and $\phi$ denote the scattering elevation and azimuth angles, respectively.  The third component is RGB color $\boldsymbol{c}(\theta,\phi)=[c_\mathrm{R}(\theta,\phi),$ $c_\mathrm{G}(\theta,\phi),c_\mathrm{B}(\theta,\phi)]^T$, which encodes scattered radiance at viewing direction $(\theta,\phi)$. Here, the subscripts R, G, and B denote the red, blue, and green components of RGB color, respectively. The relationship among these components is given by
\begin{equation}
c_i(\theta,\phi)\! = \!\int_{S^2}l_i(\theta^{\prime},\phi^{\prime})f_i(\theta,\phi,\theta^{\prime},\phi^{\prime}){\rm d}\Omega^{\prime}, i \in \{\mathrm{R},\mathrm{G},\mathrm{B}\},
\label{Bidirection GS}
\end{equation}
where $S^2$ is unit sphere, and ${\rm d}\Omega^{\prime}=\sin(\theta^{\prime}){\rm d}\theta^{\prime} {\rm d}\phi^{\prime}$. (\ref{Bidirection GS}) quantifies the angular dependence of light–object interaction. This dependence, determined by the relative configuration of the light source and the object together with the object’s material parameters, is captured by the optical scattering function $\boldsymbol{f}(\theta,\phi,\theta^{\prime},\phi^{\prime})$, which encodes the bidirectional scattering response of Gaussian ellipsoids within the environment. Next, we specify the concrete form of $\boldsymbol{f}(\theta,\phi,\theta^{\prime},\phi^{\prime})$, which is further decomposed into an angle-independent term $\boldsymbol{\rho}\in\mathbb{R}^{3}$ and an angle-dependent term $\boldsymbol{s}(\theta,\phi,\theta^{\prime},\phi^{\prime})\in\mathbb{R}^{3}$, i.e.,
\begin{equation}
\boldsymbol{f}(\theta,\phi,\theta^{\prime},\phi^{\prime})=\boldsymbol{\rho }+\boldsymbol{s}(\theta,\phi,\theta^{\prime},\phi^{\prime}), \label{Scatting function decomposition}
\end{equation}
where $\boldsymbol{\rho}$ captures the angle-independent material properties of an object (e.g., albedo).

\subsection{Optical Rendering}
Furthermore, we elaborate on the optical rendering process, which comprises two sequential stages: splatting of 3D Gaussian ellipsoids and $\alpha$-blending on the image plane. 
\subsubsection{Splatting}
After the RGB colors of all Gaussian ellipsoids are computed via (\ref{Bidirection GS}), the rendering of images initiates with splatting 3D Gaussians ellipsoids $G(\boldsymbol{x})$ onto the image plane, yielding the corresponding 2D Gaussian ellipsoids $G^{\prime}(\boldsymbol{x}_{2\text{D}})$. The splatting stage consists of two core transformations, i.e., view transformation and perspective projection. The view transformation converts the scene from absolute world coordinates to camera coordinates through an affine mapping. Subsequently, the perspective projection maps the 3D Gaussian ellipsoids onto the image plane, obtaining 2D Gaussian distributions that represent their projected forms. The mean vector and covariance matrix of one Gaussian ellipsoid after splatting are respectively expressed as
\begin{equation}
\begin{split}
&\boldsymbol{\mu}^{\prime}=\varphi(\boldsymbol{W}\boldsymbol{\mu}+\boldsymbol{d}),\\
&\boldsymbol{\Sigma}^{\prime}=\boldsymbol{J}\boldsymbol{W}\boldsymbol{\Sigma}\boldsymbol{W}^T\boldsymbol{J}^T,
\label{splatting}
\end{split}
\end{equation}
where $\boldsymbol{W}$ and $\boldsymbol{d}$ denote the rotation and translation transformation, respectively, $\boldsymbol{W}\boldsymbol{\mu}+\boldsymbol{d}$ denotes the whole view transformation, function $\varphi(\cdot): \mathbb{R}^{3} \rightarrow \mathbb{R}^{3}$ denotes the non-linear perspective projection, and $\boldsymbol{J}$ denotes the Jacobian matrix of $\varphi(\cdot)$ denoting the affine approximation of the perspective projection. 

Moreover, the 2D mean vector $\boldsymbol{\mu}_{2\text{D}}\in\mathbb{R}^{2}$ is obtained by truncating the third row of projected mean vector $\boldsymbol{\mu}^{\prime}$ and the 2D covariance matrix $\boldsymbol{\Sigma}_{2\text{D}}\in\mathbb{R}^{2\times2}$ is obtained by truncating the third row and column of the projected covariance matrix $\boldsymbol{\Sigma}^{\prime}$. Here, the perspective projection with 3D ellipsoids is to accurately characterize the transformation of objects following the optical rule that nearer objects are larger and farther objects are smaller, while the truncation is to keep consistent with the 2D image plane since the depth information has been embodied in the transformed parameters. Finally, we yield the 2D Gaussian distribution after splatting as
\begin{equation}
G^{\prime}(\boldsymbol{x}_{2\text{D}})=\exp\bigg\{-\frac{1}{2}(\boldsymbol{x}_{2\text{D}}-\boldsymbol{\mu}_{2\text{D}})^T\boldsymbol{\Sigma}^{-1}_{2\text{D}}(\boldsymbol{x}_{2\text{D}}-\boldsymbol{\mu}_{2\text{D}})\bigg\},
\label{2D_Gaussian_distribution}
\end{equation}
where $\boldsymbol{x}_{2\text{D}}$ denotes the 2D position on the image plane.
\subsubsection{$\alpha$-Blending}
Following the splatting stage, all 2D Gaussian ellipsoids are sorted according to their depth before projection (distance to the image plane). Subsequently, the optical rendering is employed to realize $\alpha$-blending, thereby computing the RGB color for each pixel on the image plane. For pixel $o$, the rendering equation is expressed as
\begin{align}
    \boldsymbol{c}^{\text{pixel}}_{o} &= \sum_{i=1}^{N} 
        \boldsymbol{c}_i(\theta,\phi) \alpha_i\prod_{k=1}^{i-1}(1-\alpha_k),
      \label{Rendering Equation New}
\end{align}
where $\alpha_i$ denotes the opacity of the $i$-th sorted ellipsoid related to the pixel, $\boldsymbol{c}_i(\theta,\phi)$ denotes ellipsoid's RGB color obtained in (\ref{Bidirection GS}), and $\boldsymbol{c}^{\text{pixel}}_o$ denotes the rendered RGB color of pixel $o$.

The opacity in (\ref{Rendering Equation New}) varies spatially between the center and the periphery of the projected 2D ellipsoid. This spatial variation is modeled by
\begin{equation}
    \alpha_{i} = \alpha^{\text{max}}_{i} G^{\prime}(\boldsymbol{x}^{\text{pixel}}_o),  \label{Gaussian_Exp}
\end{equation}
where $\alpha^{\text{max}}_i$ denotes the maximum opacity of the $i$-th ellipsoid, $\boldsymbol{x}^{\text{pixel}}_o$ represents the 2D position of pixel $o$ to be rendered on the image plane, and $G^{\prime}(\cdot)$ is defined in (\ref{2D_Gaussian_distribution}). (\ref{Gaussian_Exp}) governs the spatially varying opacity distribution within Gaussian ellipsoids, exhibiting a monotonically decreasing opacity profile with increasing radial distance from the ellipsoid center. The rendering process in (\ref{Rendering Equation New}) is completed once the colors of all pixels have been fully computed. 

\subsection{Representation and Property of Bidirectional Scattering Pattern}
\label{optical implementation detail} 
It remains to determine the bidirectional scattering component $\boldsymbol{s}(\theta,\phi,\theta^{\prime},\phi^{\prime})$ in (\ref{Scatting function decomposition}). In particular, $\boldsymbol{s}(\theta,\phi,\theta^{\prime},\phi^{\prime})$ is represented and fit by the following BSH function:
\begin{equation}
\begin{split}
\boldsymbol{s}(\theta,\phi,\theta^{\prime},\phi^{\prime})=\sum_{i=1}^{(D+1)^2}\sum_{k=1}^{(D+1)^2}\boldsymbol{a}_{ik}\boldsymbol{y}_k(\theta^{\prime},\phi^{\prime})\boldsymbol{y}_i(\theta,\phi), 
\label{BSH_BiGS}
\end{split}
\end{equation}
where $\boldsymbol{a}_{ik}\in\mathbb{R}^{3}$ denotes the learnable BSH coefficients, $D$ denotes the SH degree, $\boldsymbol{y}_i(\theta,\phi)$ denotes the $i$-th element of the SH basis vector consisting of all possible SH basis determined by the SH degree, i.e.,
\begin{equation}
\begin{split}
&\boldsymbol{y}(\theta,\phi)=\big[ y_{0,0}(\theta,\phi),y_{1,-1}(\theta,\phi),y_{1,0}(\theta,\phi),y_{1,1}(\theta,\phi)\\
&,\ldots,y_{D,-D}(\theta,\phi),\ldots,y_{D,0}(\theta,\phi),\ldots,y_{D,D}(\theta,\phi) \big],
\end{split}
\end{equation}
where $y_{n,i}(\theta,\phi)$ denotes the SH basis determined by the associated Legendre polynomials\cite{efthimiou2014spherical}, i.e.,
\begin{small}
\begin{subequations}
\begin{align}
&y_{n,i}(\theta,\phi)=\sqrt{\frac{2n+1}{2 \pi} \frac{(n-i)!}{(n+i)!}} H_{n}^{i}(\cos \theta) \cos (i\phi), i=0,1, \ldots, n, \label{Za}\\
&y_{n,i}(\theta,\phi)=\sqrt{\frac{2n+1}{2 \pi} \frac{(n-i)!}{(n+i)!}} H_{n}^{i}(\cos \theta) \sin (i\phi), i\!=\!-n, \ldots, -1, \label{Zb} \\
&H_n^i(t)=\frac{(-1)^i}{2^nn!}(1-t^2)^{i/2}\frac{d^{n+i}}{dt^{n+i}}(t^2-1)^n,\quad i=0,1,\ldots,n, \label{Zc}\\
&H_n^{-i}(t)=(-1)^i\frac{(n-i)!}{(n+i)!}H_n^i(t),\quad i=1,\ldots,n, \label{Zd}
\end{align}
\end{subequations} 
\end{small}for any $n\in \{1,\ldots,D\}$, where $H_n^i(t)$ and $H_n^{-i}(t)$ denote the associated Legendre polynomials.

Furthermore, the optical scattering function needs to satisfy the reciprocity property in order to be physically meaningful. Specifically, when the incident direction and scattering direction switch, the value of the optical scattering function should be the same. Towards this end, we enforce the bidirectional scattering component $s(\theta,\phi,\theta^{\prime},\phi^{\prime})$ to satisfy such reciprocity, which is expressed as 
\begin{equation}
\boldsymbol{s}(\theta,\phi,\theta^{\prime},\phi^{\prime})=s(\pi - \theta^{\prime},\pi + \phi^{\prime},\pi - \theta,\pi + \phi), \forall \theta,\phi,\theta^{\prime},\phi^{\prime}.
\label{Reciprocal}
\end{equation}

\subsection{Training Process}
During the training process, an adaptive density control strategy governs the number of ellipsoids and their sizes through cloning, splitting, and pruning operations. Cloning inserts new ellipsoids in under-reconstructed regions, while splitting subdivides oversized ellipsoids in over-reconstructed regions. Moreover, ellipsoids exhibiting maximum opacity below a predefined threshold undergo pruning due to their negligible contribution to the scene representation. The three adaptive density control operations are executed periodically at a fixed interval to maintain the fidelity of scene representation. Notably, the BiGS framework uses the training result of the 3DGS \cite{kerbl20233d} under static illumination as the initialization of the BiGS model, and the adaptive density control strategy is only employed during this initialization phase.

\section{BiWGS for 6D X2X CKM Construction}
\label{framework}
Motivated by BiGS that represents the radiance field via bidirectional Gaussian ellipsoids to capture dynamic illumination conditions, we present BiWGS, a novel method to construct X2X 6D CKM for arbitrary Tx-Rx position pairs.

\begin{figure}[t]
\centering
\includegraphics[width=5cm,height = 3.25cm]{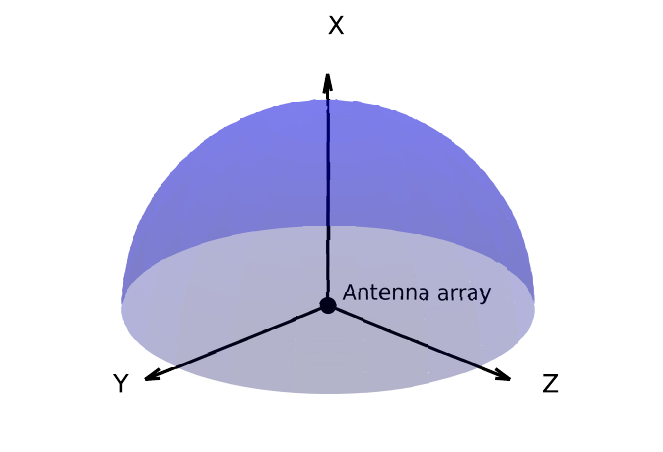}
\caption{Rx's hemispherical reception domain.}
\label{Antenna direction example}
\end{figure}

For ease of exposition, we consider a narrowband SIMO\footnote{By employing the proposed method at each transmit antenna, our framework can be easily extended to the multiple-input multiple-output (MIMO) setup with multiple antennas at both Tx and Rx.} wireless system consisting of one Tx and one Rx, in which the Tx is equipped with a single transmit antenna and the Rx is deployed with a uniform planar array (UPA) of $ N = N_v\times N_h$ antennas. It is assumed that the Rx's UPA is horizontally aligned with the ground plane. It is also assumed that the Rx's reception domain is a hemisphere shown in Fig.~\ref{Antenna direction example}, similarly as in \cite{zhao2023nerf2}.

We consider the frequency-flat fading channel model in the narrowband scenario, in which the wireless channel from Tx to Rx corresponds to the combination of multiple channel paths including reflection, scattering, and diffraction. Under this setup, we are interested in characterizing the X2X 6D CKM, which provides a function mapping from the 3D Tx-Rx position pair (a 6D parameter) to the channel knowledge. In particular, we consider the CSI of the $N\times1$ channel vector as the channel knowledge of interest. As such, the mapping via CKM is described as
\begin{equation}
    \boldsymbol{h}=q_E(\boldsymbol{p}_{\text{t}},\boldsymbol{p}_{\text{r}}),
    \label{CKM definition}
\end{equation}
where $\boldsymbol{p}_{\text{t}},\boldsymbol{p}_{\text{r}}\in \mathbb{R}^3$ denote the 3D Tx and Rx positions, respectively, the subscript $E$ denotes the wireless environment comprising both the geometric structure of the physical scene and the electromagnetic properties such as permittivity and conductivity, $\boldsymbol{h}\in \mathbb{C}^{N\times1}$ denotes the channel vector, $q_E(\cdot)$ denotes the CKM or the mapping relationship that maps from the Tx-Rx position pair to the channel vector. For simplicity, we assume that $q_E(\cdot)$ is time-invariant, which is valid for static environment or the dominant static component of time-varying environment.

In the 6D CKM construction problem, we aim to find the CKM or the mapping function $q_E(\cdot)$ based on historical channel measurements. Supposing that there are $K$ channel measurements in the training set, including the Tx-Rx position pair $\tilde{\boldsymbol{p}}_{\text{t},i},\tilde{\boldsymbol{p}}_{\text{r},i}$ and its related channel vector $\tilde{\boldsymbol{h}}_{i}$. The set of channel measurements or the training set is given by 
\begin{equation}
\begin{split}
    \mathcal{T}=\Big\{(\tilde{\boldsymbol{p}}_{\text{t},i},\tilde{\boldsymbol{p}}_{\text{r},i},\tilde{\boldsymbol{h}}_{i}) ,i=1,\dots,K\Big\}.
    \label{training set}
\end{split}
\end{equation}
The CKM construction problem is thus formulated as learning the mapping function $q_E(\cdot)$ from the training set $\mathcal{T}$.

\textit{\underline{Remark} 4.1:} In general, the mapping function $q_E(\cdot)$ of CKM can be represented in different forms depending on the adopted reconstruction methods. For instance, for interpolation-based methods, the CKM is constructed as a spatial database; for image processing-based methods, the CKM is constructed as an image; while for NeRF-based methods, the CKM is constructed as a neural network. By contrast, this paper represents the CKM via a set of bidirectional Gaussian ellipsoids.

\textit{\underline{Remark} 4.2:} Notice that conventional WRF-based designs like NeRF$^2$ \cite{zhao2023nerf2} and WRF-GS \cite{wen2025wrf} are only applicable for 3D B2X CKM with fixed $\boldsymbol{p}_{\text{r}}$, but not applicable for 6D X2X CKM in (\ref{CKM definition}) of our interest. We will propose BiWGS to construct 6D X2X CKM with varying $\boldsymbol{p}_{\text{t}}$ and $\boldsymbol{p}_{\text{r}}$ as inputs.

\begin{figure}[t]
\centering
\includegraphics[width=7.5cm,height = 4.5cm]{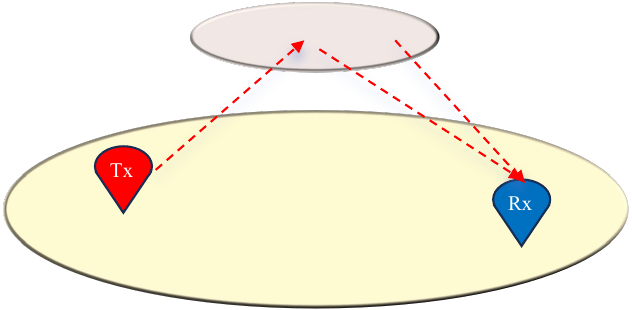}
\caption{Illustration of one ellipsoid representing a scatterer cluster. Each ellipsoid is seen as one scatterer cluster consisting of multiple scattering paths at different received angles.}
\label{cluster figure}
\end{figure}

\subsection{Scatterer Clusters based Channel Modeling and Gaussian Ellipsoid Representation}
To facilitate our BiWGS design, this subsection presents a scatterer-cluster-based channel model and represents the wireless channel through bidirectional Gaussian ellipsoids. Motivated by the widely-adopted channel models considering scatterer clusters\cite{3gpp38901}, we consider a scatterer-cluster-based channel model, which consists of a number of $S$ scatterer clusters—each contributing multiple scattering paths—and one direct path\footnote{In this paper, unless otherwise specified, arrays are assumed to be omnidirectional with unit antenna gain.}:
\begin{equation}
\scalebox{0.9}{$
\begin{split}
&\boldsymbol{h}=\sum^{S}_{i=1}\sum^{\Psi_i}_{k=1}\boldsymbol{h}^s_{i,k} + q_d\boldsymbol{h}^L\\
&=\!\underbrace{\sum^{S}_{i=1}\!\sum^{\Psi_i}_{k=1}\boldsymbol{b}(\!\theta_{i,k},\phi_{i,k}\!)
\frac{\lambda\Gamma_{i,k}(\!\theta_{i,k},\phi_{i,k},\theta_{i,k}^{\prime},\phi_{i,k}^{\prime}\!)}
{(4\pi)^{3/2} d_{\text{t},i,k}d_{\text{r},i,k}}
\Theta_{i,k} e^{-j\frac{2\pi}{\lambda}(\!d_{\text{t},i,k}\!+d_{\text{r},i,k}\!)}}_{\text{Scattering paths}}\\
&+\underbrace{q_d\boldsymbol{b}(\theta_{L},\phi_{L})
\frac{\lambda}{4\pi d_{L}} \Theta_Le^{-j\frac{2\pi}{\lambda}d_L}}_{\text{Direct path}}.
\end{split}
$}
\label{General Channel model}
\end{equation}
In (\ref{General Channel model}), $\Psi_i$ denotes the number of scattering paths of each scatterer cluster $i$, $\boldsymbol{h}^s_{i,k}$ denotes the channel vector of the $(i,k)$-th scattering path, $\boldsymbol{h}^L$ denotes the channel vector of direct channel path, $\lambda$ denotes the wavelength, $\theta_{i,k}$, $\phi_{i,k}$, $\theta_{i,k}^{\prime}$, and  $\phi_{i,k}^{\prime}$ denote the scattering elevation angle, scattering azimuth angle, incident elevation angle, and incident azimuth angle of the $(i,k)$-th scattering path, respectively, $d_{\text{t},i,k}$ and $d_{\text{r},i,k}$ denote the distance to Tx and Rx in the $(i,k)$-th scattering path, respectively, $\Theta_{i,k}\in\mathbb{C}$ denotes the complex attenuation coefficient representing extra amplitude attenuation and phase shifting induced by the obstruction along the $(i,k)$-th scattering path. Furthermore, $\Gamma(\theta_{i,k},\phi_{i,k},\theta^{\prime}_{i,k},\phi^{\prime}_{i,k}) \in \mathbb{C}$ denotes the bidirectional complex scattering coefficient, denoting the change of amplitude and phase caused by scattering. Notably, $\Gamma(\theta_{i,k},\phi_{i,k},\theta^{\prime}_{i,k},\phi^{\prime}_{i,k})$ is fundamentally related to bi-static radar cross section, which characterizes the bidirectional angular dependence of scattering on both incident and scattering angles\cite{huang2021anisotropic}. For the direct path, $\Theta_L\in\mathbb{C}$ is a complex attenuation coefficient caused by obstruction along the direct path, $\theta_L$ and $\phi_L$ denote the elevation and azimuth angles of the direct path, respectively, $d_L$ denotes the distance between Tx and Rx. Notably, the LOS path is a special case of the direct path without any obstruction. $q_d$ is a 0-1 indicator denoting if a direct path exists (Tx lies within the Rx's hemispherical reception domain, as illustrated in Fig.~\ref{Antenna direction example}), which is given by
\begin{equation}
q_d = 
\begin{cases}
0, \quad \text{Tx is not within the Rx's reception domain},\\
1,  \quad \text{Tx is within the Rx's reception domain}.
\label{qd}
\end{cases}
\end{equation}
In addition, $\boldsymbol{b}(\cdot)$ denotes the steering vector. For the UPA, we have
\begin{equation}
\begin{split}
&\boldsymbol{b}_v(\theta_{i,k})=\frac{1}{N_v}[1,e^{j2\pi\frac{d_v}{\lambda}\!\cos\theta_{i,k}},\dots,e^{j2\pi\frac{d_v}{\lambda}(N_v-1)\!\cos\theta_{i,k}}]^T, \\
&\boldsymbol{b}_h(\theta_{i,k},\phi_{i,k})\\
&\!\!=\!\!\frac{1}{N_h}[1,e^{j2\pi\frac{d_h}{\lambda}\!\sin\theta_{i,k} \!\sin\phi_{i,k}},\dots,e^{j2\pi\frac{d_h}{\lambda}(N_h-1)\!\sin\theta_{i,k} \!\sin\phi_{i,k}}]^T,\\
&\boldsymbol{b}(\theta_{i,k},\phi_{i,k}) = \boldsymbol{b}_v(\theta_{i,k}) \otimes \boldsymbol{b}_h(\theta_{i,k},\phi_{i,k}),
\label{steering vector}
\end{split}
\end{equation}
where $d_h$ and $d_v$ represent the horizontal and vertical spacing between two adjacent antennas, respectively.

Note that in (\ref{General Channel model}), $S$ is a parameter depending on the environment. The reflection and diffraction can be equivalently represented by a cluster of scatterers. Furthermore, in (\ref{General Channel model}), we only consider the case with one single hop of scattering by ignoring the multi-hop scattering. This is consistent with channel models in \cite{huang2023joint,zhang2025efficient,chowdary2024hybrid} to facilitate our proposed BiWGS design\footnote{Though we design BiWGS based on the simplified model with single-hop scattering, it will be shown in Section \ref{results} that the proposed BiWGS can well represent the practical wireless environment with multi-hop transmission paths considered.}.

In the BiWGS design, we employ Gaussian ellipsoids as virtual scatterer clusters corresponding to one or more scattering paths\cite{zhang2025unified}, as shown in Fig.~\ref{cluster figure}. To represent the wireless channel via Gaussian ellipsoids, we first reformulate the channel model in (\ref{General Channel model}), by expressing it as the combination of scattering signal paths from every possible AoA of Rx, i.e.,
\begin{equation}
\begin{split}
\boldsymbol{h}
&=\sum^{S}_{i=1}\sum^{\Psi_i}_{k=1}\boldsymbol{h}^s_{i,k} + q_d\boldsymbol{h}^L\\
&=\int^{\pi/2}_{-\pi/2}\int^{\pi}_{0}\boldsymbol{h}^s(\theta,\phi) {\rm d}\theta {\rm d}\phi + q_d \boldsymbol{h}^L,\\
&\approx\sum_{\phi \in \mathcal{Z}}\sum_{\theta \in \mathcal{V}} \boldsymbol{h}^s(\theta,\phi) + q_d \boldsymbol{h}^L,
\label{Equivalent Channel model}
\end{split}
\end{equation}
where $\boldsymbol{h}^s(\theta,\phi)$ denotes the scattering channel vector related to the AOA $(\theta,\phi)$, and $\mathcal{Z}$ and $\mathcal{V}$ denote the set of sampled azimuth and elevation angles, respectively, by discretization of AOAs at the Rx with a specified angular resolution (determined by the Rx's reception domain in Fig.~\ref{Antenna direction example}). Note that for each sampled ray along the angle $(\theta,\phi)$, it may penetrate multiple Gaussian ellipsoids, each of which results in a distinct scattering path associated with that ellipsoid. As such, multiple rays may traverse the same ellipsoid to generate multiple resolvable scattering paths. Let $\mathcal{E}_{\text{r}}(\theta,\phi)$ denote the set of ellipsoids that are penetrated by the sampled ray related to AOA $(\theta,\phi)$. The channel model in (\ref{General Channel model}) is reformulated as
\begin{equation}
\begin{split}
\boldsymbol{h}
&\approx\sum_{\phi \in \mathcal{Z}}\sum_{\theta \in \mathcal{V}}\sum_{m \in \mathcal{E}_{\text{r}}(\theta,\phi)} \bigg[ \boldsymbol{b}(\theta,\phi)\frac{\lambda\Gamma_m(\theta_m,\phi_m,\theta_m^{\prime},\phi_m^{\prime})}{(4\pi)^{3/2} d_{\text{t},m}d_{\text{r},m}}\\
&\cdot \Theta_{m}(\theta,\phi,\boldsymbol{p}_{\text{t}},\boldsymbol{p}_{\text{r}},\boldsymbol{p}_m) e^{-j\frac{2\pi}{\lambda}(d_{\text{t},m}+d_{\text{r},m})} \bigg]  \\
&+q_d\boldsymbol{b}(\theta_{L},\phi_{L})\frac{\lambda}{4\pi d_{L}} \Theta_L(\boldsymbol{p}_{\text{t}},\boldsymbol{p}_{\text{r}})e^{-j\frac{2\pi}{\lambda}d_L},
\label{Reformulated Channel model}
\end{split}
\end{equation}
where $\boldsymbol{p}_m$ denotes the position of ellipsoid $m$, $\Theta_{m}(\theta,\phi,\boldsymbol{p}_{\text{t}},\boldsymbol{p}_{\text{r}},\boldsymbol{p}_m)$ denotes the complex attenuation, which is determined by the AOA, position of ellipsoid $m$, Tx position, and Rx position. Under our Gaussian ellipsoid representation, the attenuation $\Theta_{m}(\theta,\phi,\boldsymbol{p}_{\text{t}},\boldsymbol{p}_{\text{r}},\boldsymbol{p}_m)$ is caused by the obstruction related to ellipsoid $m$ at AOA ($\theta,\phi$) of Rx.

In particular, we define the 3D Gaussian ellipsoids similarly as in (\ref{Gaussian_distribution}) and (\ref{covariance}), which capture both scattering behavior and obstruction attenuation during the signal transmission. In the following, we use Gaussian ellipsoids to represent the complex attenuation and bidirectional complex scattering coefficient. Specifically, $\Theta_{m}(\theta,\phi,\boldsymbol{p}_{\text{t}},\boldsymbol{p}_{\text{r}},\boldsymbol{p}_m)$ will be determined by wireless splatting and wireless rendering, and $\Gamma_m(\theta_m,\phi_m,\theta_m^{\prime},\phi_m^{\prime})$ will be represented by BSH function, as will be elaborated in detail next.

\begin{figure}[t]
\centering
\includegraphics[width=\columnwidth]{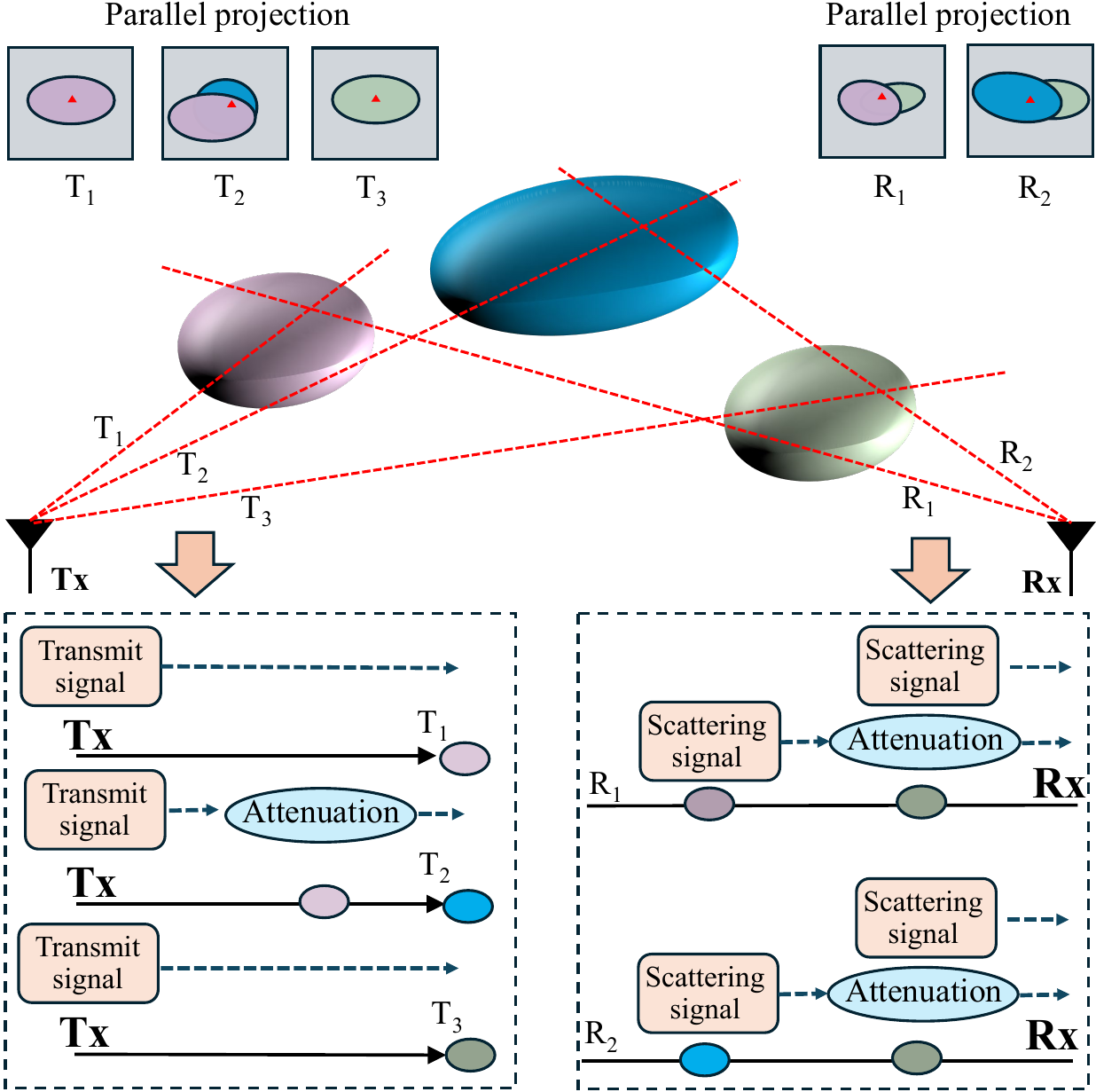}
\caption{Illustration of Tx-side and Rx-side wireless splatting. T$_1$, T$_2$, and T$_3$ represent three Tx-ellipsoid paths associated with three ellipsoids, respectively. R$_1$ and R$_2$ represent two different sampled rays of Rx, respectively.}
\label{projection illustration}
\end{figure}

\subsection{Representation of $\Theta_{m}(\theta,\phi, \textbf{p}_{\rm t},\textbf{p}_{\rm r},\textbf{p}_m)$ via Wireless Splatting and Rendering}
This subsection delineates the representation of complex attenuation $\Theta_{m}(\theta,\phi,\boldsymbol{p}_{\text{t}},\boldsymbol{p}_{\text{r}},\boldsymbol{p}_m)$ via wireless splatting and wireless rendering process within the BiWGS framework. 

\subsubsection{Wireless Splatting}
\label{wireless splatting section}
The wireless splatting stage entails two geometric transformations: a view transformation and a parallel projection. Different from conventional splatting in (\ref{splatting}) operated on an image plane defined by a physical camera's pose, in this paper, we use virtual projection planes to realize wireless splatting. These virtual projection planes are established by specifying a normal vector and an anchor point in 3D space, which uniquely defines the plane's orientation and position. Notably, the wireless splatting is different on the Tx and Rx side due to different choices of normal vectors. 

In the Tx-side wireless splatting, each ellipsoid will receive the incident signal only from the Tx due to the consideration of single-hop scattering only. Therefore, we consider only one virtual projection plane. For ellipsoid $m$, the anchor point is the position of the ellipsoid itself, and the normal vector is chosen as the incident unit direction vector, which is expressed as
\begin{equation}
\begin{split}
&\boldsymbol{n}_{\text{t}}= \frac{\boldsymbol{p}_m-\boldsymbol{p}_{\text{t}}}{\Vert \boldsymbol{p}_m-\boldsymbol{p}_{\text{t}}\Vert_2}.
\end{split}
\end{equation}

In the Rx-side wireless splatting, multiple virtual projection planes are established for different sampled AOAs, each corresponding to a distinct transmission path. The anchor point for all planes is set at the position of Rx. For certain AOA $(\theta_o,\phi_o)$, the normal vector of virtual projection plane is expressed as 
\begin{equation}
    \boldsymbol{n}_{\text{r}}=[\sin(\theta_o)\cos(\phi_o),\sin(\theta_o)\sin(\phi_o),\cos(\theta_o)]^T.
\end{equation}
Fig.~\ref{projection illustration} shows both the Tx-side and Rx-side wireless splatting.

\begin{figure}[t]
    \centering
    \subfloat[Parallel projection.]{
        \includegraphics[width=3.25cm,height=3.5cm]{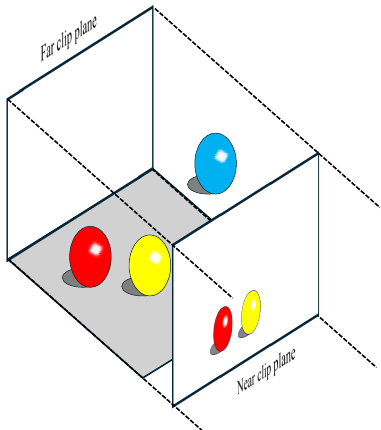}
        \label{parallel projection}
    }
    \hspace{0.1\linewidth} 
    \subfloat[Perspective projection.]{
        \includegraphics[width=3.25cm,height=3.5cm]{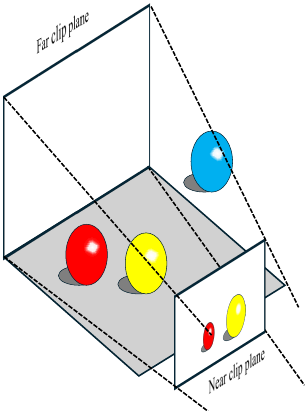}
        \label{perspective projection}
    }
    \caption{Comparison of projection methods: (a) Parallel projection for CKM construction of our interest. (b) Perspective projection in 3DGS and BiGS.}
    \label{projection comparison figure}
\end{figure}

However, the implementation of the parallel projection differs significantly from its perspective projection counterpart in (\ref{splatting}). In the computer graphics domain, perspective projection is used to emulate human vision or camera image acquisition by simulating depth perception. By contrast, our methodology employs parallel projection, which maintains distance-invariant projection scaling. This distinction is caused by the fact that there is no camera in a wireless communication context, thereby eliminating the requirement for a perspective projection. A comparison of the two above projection methods is shown in Fig.~\ref{projection comparison figure}. 
 
After the virtual projection plane is established, the wireless splatting is expressed as
\begin{equation}
\begin{split}
\boldsymbol{\mu}^{\prime}_{\text{t}}&=\boldsymbol{W}_{\text{t}}\boldsymbol{\mu}+\boldsymbol{d}_{\text{t}}=[\mu^{\prime}_{\text{t},x},\mu^{\prime}_{\text{t},y},\mu^{\prime}_{\text{t},z}]^T,\\
\boldsymbol{\Sigma}^{\prime}_{\text{t}}&=\boldsymbol{W}_{\text{t}}\boldsymbol{\Sigma}\boldsymbol{W}^T_{\text{t}},\\
\boldsymbol{\mu}^{\prime}_{\text{r}}&=\boldsymbol{W}_{\text{r}}\boldsymbol{\mu}+\boldsymbol{d}_{\text{r}}=[\mu^{\prime}_{\text{r},x},\mu^{\prime}_{\text{r},y},\mu^{\prime}_{\text{r},z}]^T,\\
\boldsymbol{\Sigma}^{\prime}_{\text{r}}&=\boldsymbol{W}_{\text{r}}\boldsymbol{\Sigma}\boldsymbol{W}^T_{\text{r}}.
\label{wireless splatting}
\end{split}
\end{equation}
Notably, the distances $d_{\text{t}, m}$ and $d_{\text{r}, m}$ in (\ref{Reformulated Channel model}) are also determined during the wireless splatting process. Based on (\ref{wireless splatting}), the distances are expressed as
\begin{equation}
\begin{split}
    d_{\text{t}, m} = \mu^{\prime}_{\text{t}z},\quad
    d_{\text{r}, m} = \mu^{\prime}_{\text{r}z}.
    \label{distances}
\end{split}
\end{equation}

Subsequently, we truncate the third row of mean vectors $\boldsymbol{\mu}^{\prime}_{\text{t}}$, $\boldsymbol{\mu}^{\prime}_{\text{r}}$ and third row/column of covariance matrices $\boldsymbol{\Sigma}^{\prime}_{\text{t}}$, $\boldsymbol{\Sigma}^{\prime}_{\text{r}}$. These operations yield the 2D Gaussian parameters $\boldsymbol{\mu}_{2\text{D}}$ and $\boldsymbol{\Sigma}_{2\text{D}}$ for projected 2D Gaussian ellipsoids. The 2D Gaussian distributions after wireless splatting for Tx and Rx sides are denoted as $G^{\prime}_{\text{t}}(\cdot)$ and $G^{\prime}_{\text{r}}(\cdot)$, respectively. The concrete form is the same as that in  (\ref{2D_Gaussian_distribution}).

\subsubsection{Wireless Rendering}
After the wireless splatting stage, the complex attenuation caused by obstruction among ellipsoids is calculated by the wireless rendering equation. For an ellipsoid $m$, the complex obstruction attenuation of its relevant scattering path is decomposed into Tx-side attenuation $\Theta_{\text{t}, m}(\boldsymbol{p}_{\text{t}},\boldsymbol{p}_m)$, and Rx-side attenuation $\Theta_{\text{r}, m}(\theta,\phi,\boldsymbol{p}_{\text{r}},\boldsymbol{p}_m)$, which is given by
\begin{equation}
\Theta_{m}(\theta,\phi,\boldsymbol{p}_{\text{t}},\boldsymbol{p}_{\text{r}},\boldsymbol{p}_m) = \Theta_{\text{t}, m}(\boldsymbol{p}_{\text{t}},\boldsymbol{p}_m)\Theta_{\text{r}, m}(\theta,\phi,\boldsymbol{p}_{\text{r}},\boldsymbol{p}_m).
\label{Decomposition}
\end{equation}

For the Tx-side wireless rendering, the wireless rendering equation is given by
\begin{equation}
\Theta_{\text{t}, m}(\boldsymbol{p}_{\text{t}},\boldsymbol{p}_m)= \prod_{k\in \mathcal{F}_{\text{t}}(m)}(1-\alpha_{m,k})e^{-j\frac{2\pi}{\lambda}\gamma_{m,k}},
\label{wireless rendering equation_Tx}
\end{equation}
where $\mathcal{F}_{\text{t}}(m)$ denotes the set of ellipsoids along the Tx-ellipsoid path, which is determined in the wireless splatting stage, $\alpha_{m,k}$ denotes the opacity of ellipsoid $k$, $\gamma_{m,k}\in[0,\lambda]$ denotes the length of equivalent path that signal travels within ellipsoid $k$, physically related the refractive index of material, and $(1-\alpha_{m,k})e^{-j\frac{2\pi}{\lambda}\gamma_{m,k}}$ represents the complex attenuation caused by the obstruction of ellipsoid $k$. 

In (\ref{wireless rendering equation_Tx}), $\alpha_{m,k}$ and $\gamma_{m,k}$ are obtained in the same way similarly as in (\ref{Gaussian_Exp}) based on the splatted 2D Gaussian ellipsoids, which is given by
\begin{equation}
\begin{split}
    &\alpha_{m,k} = \alpha^{\text{max}}_{k} G^{\prime}_{\text{t},m,k}(\boldsymbol{x}^{m}_{2\text{D}}),\\
    &\gamma_{m,k} = \gamma^{\text{max}}_{k} G^{\prime}_{\text{t},m,k}(\boldsymbol{x}^{m}_{2\text{D}}),
    \label{Wireless Gaussian_Exp_Tx}
\end{split}
\end{equation}
where $\alpha^{\text{max}}_{k}$ and $\gamma^{\text{max}}_{k}$ denote the maximum opacity and length of equivalent path of ellipsoid $k$, respectively, $G^{\prime}_{\text{t},m,k}(\cdot)$ denotes the 2D Gaussian distribution of ellipsoid $k$ via the Tx-side wireless splatting of ellipsoid $m$, and $\boldsymbol{x}^{m}_{2\text{D}}$ denotes the 2D position of ellipsoid $m$ in the virtual projection plane.

Furthermore, for the Rx-side rendering, the wireless rendering equation is given by

\begin{equation}
\Theta_{\text{r}, m}(\theta,\phi,\boldsymbol{p}_{\text{r}},\boldsymbol{p}_m)= \alpha_{m,m}\prod_{l\in \mathcal{F}_\text{r}(m)}(1-\alpha_{m,l})e^{-j\frac{2\pi}{\lambda}\gamma_{m,l}},
\label{wireless rendering equation_Rx}
\end{equation}
where $\mathcal{F}_\text{r}(m)$ denotes the set of other ellipsoids appearing along the ellipsoid-Rx path obstructing the transmission, which is also determined in the wireless splatting stage. $\alpha_{m,m}$, $\alpha_{m,l}$ and $\gamma_{m,l}$ are defined similarly as in (\ref{Wireless Gaussian_Exp_Tx}):
\begin{equation}
\begin{split}
    \alpha_{m,m} &= \alpha^{\text{max}}_{m} G^{\prime}_{\text{r},m,m}(\boldsymbol{x}^{\text{r}}_{2\text{D}}),\\
    \alpha_{m,l} &= \alpha^{\text{max}}_{l} G^{\prime}_{\text{r},m,l}(\boldsymbol{x}^{\text{r}}_{2\text{D}}),\\
    \gamma_{m,l} &= \gamma^{\text{max}}_{l} G^{\prime}_{\text{r},m,l}(\boldsymbol{x}^{\text{r}}_{2\text{D}}),
    \label{Wireless Gaussian_Exp_Rx}
\end{split}
\end{equation}
where $G^{\prime}_{\text{r},m,l}(\cdot)$ denotes the 2D Gaussian distribution of ellipsoid $l$ via the Rx-side wireless splatting of ellipsoid $m$, and $\boldsymbol{x}^{r}_{2\text{D}}$ denotes the 2D position of Rx in the virtual projection plane. 

Note that the discrepancy between (\ref{Wireless Gaussian_Exp_Tx}) and (\ref{Wireless Gaussian_Exp_Rx}) stems from the different anchor point chosen in the Tx-side and Rx-side wireless splatting. Also note that the wireless rendering in (\ref{wireless rendering equation_Tx}) and (\ref{wireless rendering equation_Rx}) is different from the optical rendering counterpart in (\ref{Rendering Equation New}). Different from (\ref{Rendering Equation New}) focusing on real RGB values, the formulations in (\ref{wireless rendering equation_Tx}) and (\ref{wireless rendering equation_Rx}) incorporate two critical physical mechanisms: amplitude attenuation governed by the Gaussian ellipsoid's opacity $\alpha$, which quantifies signal amplitude reduction along obstructed transmission paths, and phase shifting introduced through the length of equivalent path $\gamma$, capturing phase distortion due to dielectric interactions in obstacles. This extends traditional rendering theory to scenarios of wireless signal transmission.

In addition, we obtain the complex obstruction attenuation $\Theta_L(\boldsymbol{p}_{\text{t}},\boldsymbol{p}_{\text{r}})$, similarly as the Tx-side wireless rendering procedure, by employing sequential direct wireless splatting and rendering operations. However, there are notable differences. First, for wireless splatting of direct path, the anchor point is the Rx position, and the normal vector for virtual projection plane is given as
\begin{equation}
    \boldsymbol{n}_{d}=\frac{\boldsymbol{p}_{\text{r}}-\boldsymbol{p}_{\text{t}}}{\Vert \boldsymbol{p}_{\text{r}}-\boldsymbol{p}_{\text{t}}\Vert_2}.
\end{equation}
Moreover, the distance of the direct path $d_L$ is equal to the distance between Tx and Rx. As such, the complex attenuation $\Theta_L$ of direct path is expressed as
\begin{equation}
\begin{split}
\Theta_L(\boldsymbol{p}_{\text{t}},\boldsymbol{p}_{\text{r}})=\prod_{k\in \mathcal{E}_\text{r}(\theta_L,\phi_L)}(1-\alpha_{k})e^{-j\frac{2\pi}{\lambda}\gamma_{k}},
\label{LOS rendering}
\end{split}
\end{equation}
where $\mathcal{E}_\text{r}(\theta_L,\phi_L)$ denotes the set of ellipsoids along the direct path.

\textit{\underline{Remark} 4.3:} Comparing the wireless rendering equations (\ref{wireless rendering equation_Tx}), (\ref{wireless rendering equation_Rx}), and (\ref{LOS rendering}) with the optical rendering equation (\ref{Rendering Equation New}) highlights a decisive difference: phase. The wireless formulation retains phase, whereas the optical counterpart discards it because visible light operates at much higher frequencies (hence much shorter wavelengths), causing phase to oscillate beyond the temporal resolution of standard detectors and be hard to trace\cite{zhao2023nerf2}. Nevertheless, phase is important in wireless communication systems due to its critical role in constructive/destructive signal combination of multipath components, which directly governs signal integrity at Rx. 

\textit{\underline{Remark} 4.4:} Another distinction between optical rendering and wireless transmission models lies in the absence of distance-dependent attenuation terms in the optical rendering equation. This distinction arises from fundamental differences in receiver perception mechanisms. Distance-dependent attenuation occurs in both wireless communications and free-space optical (FSO) systems\cite{kaushal2017free}, causing a reduction of absolute intensity. Crucially, these systems employ electronic receivers capable of measuring the absolute intensity of EM wave. Nevertheless, within the computer graphics field (including 3DGS and BiGS), rendered images are ultimately perceived by the human eyes. Human eyes employ complicated physiological filtering and processing on incident light, and perceive relative intensity of light in a logarithmic way\cite{gross2008human}. Therefore, distance-dependent attenuation is negligible for human eyes except at sufficiently large distances, which is rarely encountered in optical 3D reconstruction contexts. 

\subsection{Representation of Bidirectional Complex Scattering Coefficient $\Gamma_m(\theta_m,\phi_m,\theta_m^{\prime},\phi_m^{\prime})$ via BSH}
This subsection delineates the representation of bidirectional complex scattering coefficient $\Gamma_m(\theta_m,\phi_m,\theta_m^{\prime},\phi_m^{\prime})$ via BSH within the BiWGS framework. Similar to BiGS, we employ the BSH function to fit the bidirectional complex scattering coefficient. Mathematically, BSH can directly approximate both real and imaginary components of $\Gamma_m(\theta_m,\phi_m,\theta_m^{\prime},\phi_m^{\prime})$. However, experimentally, we find that directly fitting the bidirectional complex scattering coefficient via BSH may easily lead to training instability, manifesting as gradient explosion. To stabilize the convergence, we decompose $\Gamma_m(\theta_m,\phi_m,\theta_m^{\prime},\phi_m^{\prime})$ into two coefficients to optimize separately during the training process, which is given by
\begin{equation}
    \Gamma_m(\theta_m,\phi_m,\theta_m^{\prime},\phi_m^{\prime}) = \underbrace{Z_m}_{\text{Angle-independent}}
    \underbrace{V_m(\theta_m,\phi_m,\theta_m^{\prime},\phi_m^{\prime}) }_{\text{Angle-dependent}},
    \label{training trick}
\end{equation}
where the coefficient $Z_m\in\mathbb{R}$ denotes an angle-independent coefficient that will be directly optimized in the backward propagation, and $V_m(\theta_m,\phi_m,\theta_m^{\prime},\phi_m^{\prime})$ denotes an angle-dependent coefficient satisfying $\lvert V_m(\theta_m,\phi_m,\theta_m^{\prime},\phi_m^{\prime})\rvert\leq1$ that will be fit by the BSH function. With the decomposition in (\ref{training trick}), the training of the proposed BiWGS method will become more stable.

Next, the coefficient $V_m(\theta_m,\phi_m,\theta_m^{\prime},\phi_m^{\prime})$ is fit via the BSH function. To facilitate the understanding, the coefficient is rewritten as
\begin{equation}
\begin{split}
&V_{m}(\theta_m,\phi_m,\theta_m^{\prime},\phi_m^{\prime})\\
&=\mathcal{R}(V_{m}(\theta_m,\phi_m,\theta_m^{\prime},\phi_m^{\prime}))+j\mathcal{I}(V_{m}(\theta_m,\phi_m,\theta_m^{\prime},\phi_m^{\prime})),
\end{split}
\end{equation}
where $\mathcal{R}(\cdot)$ and $\mathcal{I}(\cdot)$ denote the real and imaginary parts of a complex number, respectively. Next, we use 2 groups of BSH coefficients to fit the real and imaginary parts of $V_{m}(\theta_m,\phi_m,\theta_m^{\prime},\phi_m^{\prime})$, respectively, which are expressed as
\begin{equation}
\begin{split}
&\mathcal{R}(V_{m}(\theta_m,\phi_m,\theta_m^{\prime},\phi_m^{\prime}))\\
&=\sum_{i=1}^{(D+1)^2}\sum_{k=1}^{(D+1)^2}a^{\text{Re}}_{i,k,m}\boldsymbol{y}_k(\theta^{\prime}_m,\phi^{\prime}_m)\boldsymbol{y}_i(\theta_m,\phi_m) \label{Re_SH},
\end{split}
\end{equation}
\begin{equation}
\begin{split}
&\mathcal{I}(V_{m}(\theta_m,\phi_m,\theta_m^{\prime},\phi_m^{\prime}))\\
&=\sum_{i=1}^{(D+1)^2}\sum_{k=1}^{(D+1)^2}a^{\text{Im}}_{i,k,m}\boldsymbol{y}_k(\theta^{\prime}_m,\phi^{\prime}_m)\boldsymbol{y}_i(\theta_m,\phi_m) \label{Im_SH},
\end{split}
\end{equation}
where $a^{\text{Re}}_{i,k,m}$ and $a^{\text{Im}}_{i,k,m}$ are BSH coefficients of the real part and imaginary part, respectively.

Note that $V_m(\theta,\phi,\theta^{\prime},\phi^{\prime})$ should also fit the reciprocity, similarly as that in (\ref{Reciprocal}), in which the bidirectional complex scattering coefficient remains identical when the incident and scattering directions are interchanged. In other words, we have
\begin{equation}
V_m(\theta,\phi,\theta^{\prime},\phi^{\prime})\! =\! V_m(\pi - \theta^{\prime}\!,\!\pi + \phi^{\prime}\!,\!\pi - \theta\!,\!\pi + \phi),\forall \theta,\phi,\theta^{\prime},\phi^{\prime}.
\label{wireless Reciprocal}
\end{equation}
To ensure the reciprocity condition in (\ref{wireless Reciprocal}), we impose the symmetric structures on the BSH coefficients at different indexes $i,k$ in (\ref{Re_SH}) and (\ref{Im_SH}). Towards this end, when $D=3$, we define two index sets partitioning the SH basis according to their degrees for clarity\footnote{These two index sets are defined based on the definition and properity of SH basis. More mathematical details can be found in \cite{efthimiou2014spherical,ramamoorthi2004signal}.} expressed as
\begin{equation}
\begin{split}
&\mathcal{D}_e=\{1,5,6,7,8,9\},\\
&\mathcal{D}_o=\{2,3,4,10,11,12,13,14,15,16\}.
\end{split}
\end{equation}
According to the concrete forms of the SH basis (\ref{Za}), (\ref{Zb}), the following properties are obtained
\begin{equation}
\begin{cases}
\boldsymbol{y}_i(\theta_m,\phi_m) = \boldsymbol{y}_i(\pi-\theta_m,\pi+\phi_m),  \quad i\in\mathcal{D}_e,\\
\boldsymbol{y}_i(\theta_m,\phi_m) = -\boldsymbol{y}_i(\pi-\theta_m,\pi+\phi_m),  \quad i\in\mathcal{D}_o.
\label{sh basis property}
\end{cases}
\end{equation}
Substituting (\ref{Re_SH}), (\ref{Im_SH}), and (\ref{sh basis property}) into (\ref{wireless Reciprocal}), we obtain
\begin{equation}
\begin{cases}
a_{i,k,m}^l=a_{k,i,m}^l,  \quad i,k\in\mathcal{D}_e \lor i,k\in\mathcal{D}_o,  \\
a_{i,k,m}^l=-a_{k,i,m}^l,  \quad \text{otherwise},
\label{sh coefficient property}
\end{cases}
\end{equation}
for any $l \in \{\text{Re},\text{Im}\}$. By preserving the symmetric structures of BSH coefficients specified in (\ref{sh coefficient property}), reciprocity is guaranteed.

It is also interesting to compare the BSH representation in BiGS for optical rendering versus our proposed BiWGS for CKM construction. The BiGS method employs the 3 groups of BSH coefficients for each ellipsoid to model the optical bidirectional scattering function, thereby determining the ellipsoid's RGB color attributes for specific viewing directions under certain illumination. In contrast, the proposed BiWGS method adapts 2 groups of BSH coefficients to characterize bidirectional complex scattering coefficients for each ellipsoid, thus fitting the complex scattering pattern for specific Tx-Rx position pairs.

\begin{figure*}[t!]
\centering
\includegraphics[width=0.9\linewidth]{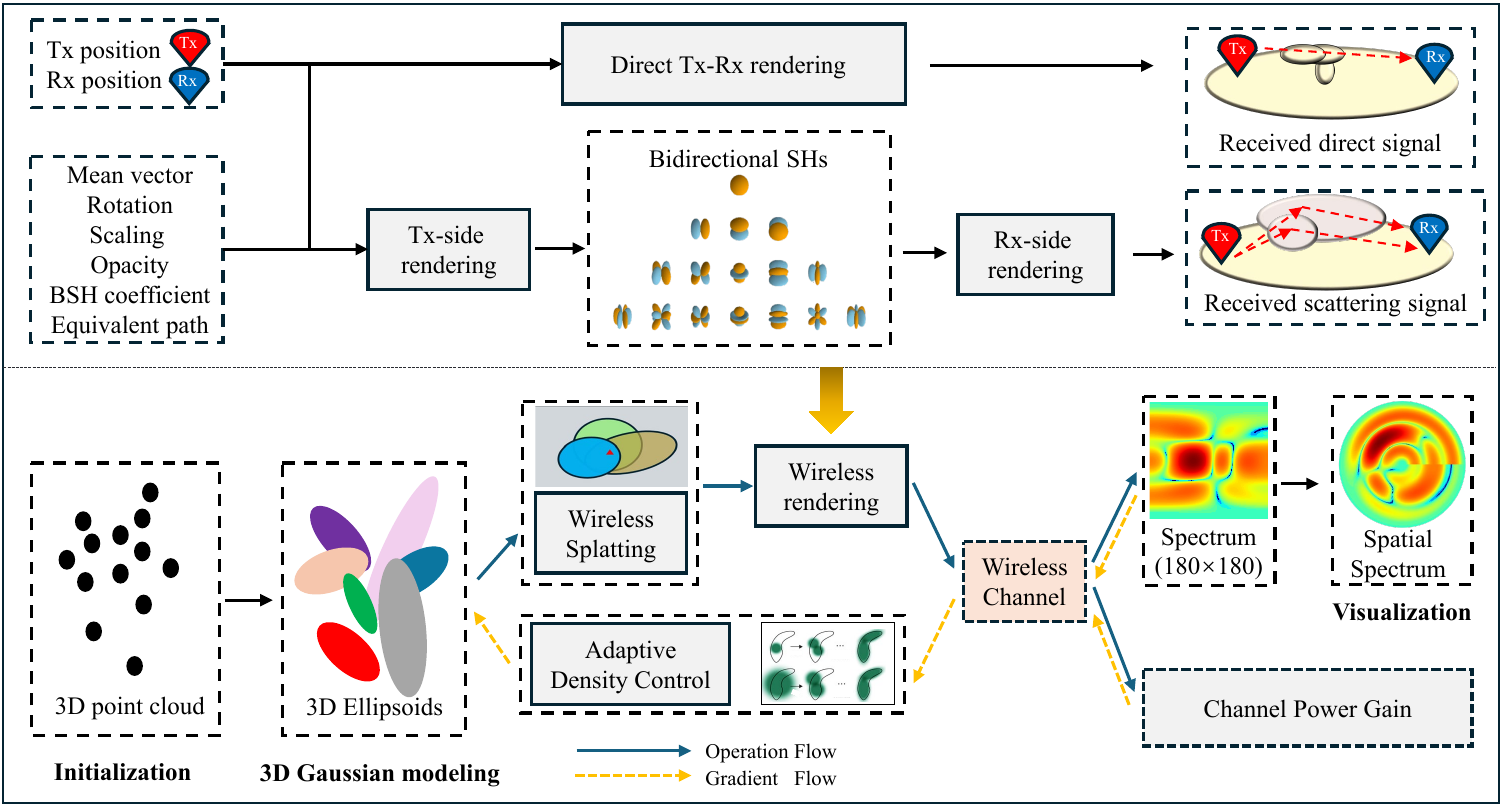}
\caption{Illustration of BiWGS.}
\label{pipeline}
\end{figure*}

\subsection{Overall Process}
\label{training process}

Substituting (\ref{wireless rendering equation_Tx}), (\ref{wireless rendering equation_Rx}), (\ref{LOS rendering}), and (\ref{training trick}) into (\ref{Reformulated Channel model}), we obtain the wireless channel represented by BiWGS method as
\begin{equation}
\begin{split}
&\boldsymbol{h} \approx\sum_{\phi \in \mathcal{Z}}\sum_{\theta \in \mathcal{V}} \sum_{m\in\mathcal{E}_{\text{r}}(\theta,\phi) } \Bigg[ \boldsymbol{b}(\theta,\phi)\\
&\cdot \frac{e^{-j\frac{2\pi}{\lambda}d_{\text{t},m}}}{\sqrt{4\pi}d_{\text{t},m}} \underbrace{\prod_{k\in \mathcal{F}_\text{t}(m)}(1-\alpha_{m,k})e^{-j\frac{2\pi}{\lambda}\gamma_{m,k} }}_{\text{Tx-side wireless rendering}}\\
&\cdot \underbrace{Z_mV_{m}(\theta_m,\phi_m,\theta^{\prime}_m,\phi^{\prime}_m)}_{\text{Bidirectional complex scattering coefficient}}\\
&\cdot \frac{\lambda e^{-j\frac{2\pi}{\lambda}d_{\text{r},m}}}{4\pi d_{\text{r},m}}  \underbrace{  \alpha_{m,m}\prod_{l \in \mathcal{F}_\text{r}(m)}(1-\alpha_{m,l})e^{-j\frac{2\pi}{\lambda}\gamma_{m,l}}}_{\text{Rx-side wireless rendering}}\Bigg]\\
&+q_d\boldsymbol{b}(\theta_L,\phi_L)\frac{\lambda e^{-j\frac{2\pi}{\lambda}d_L}}{4\pi d_{L}}\underbrace{\prod_{k\in \mathcal{E}_{\text{r}}(\theta_L,\phi_L)}(1-\alpha_{k})e^{-j\frac{2\pi}{\lambda}\gamma_{k}}}_{\text{Direct path wireless rendering}}.
\label{total wireless channel}
\end{split}
\end{equation}

\begin{table*}[t]
\centering
\caption{Major Differences: BiGS versus BiWGS}
\begin{tabular}{lcc}
\toprule
\textbf{Characteristic} & \textbf{BiGS} & \textbf{BiWGS} \\
\midrule
Task & 3D reconstruction under dynamic illumination & 6D CKM construction under varying Tx-Rx position pair\\
Target of rendering & Real RGB color $\boldsymbol{c}^{\text{pixel}}_o$ of each pixel of image & Complex channel vector $\boldsymbol{h}$ \\
Projection method & Perspective projection & Parallel projection\\
Distance-dependent attenuation & Not explicitly modeled & Explicitly modeled \\
Number of projection planes & Single image plane at camera & Multiple virtual projection planes at Rx \\
Phase information & Not modeled & Essential component\\
BSH fitting objective  & Optical scattering function $\boldsymbol{f}(\theta,\phi,\theta^{\prime},\phi^{\prime})$ & Bidirectional complex scattering coefficient $\Gamma(\theta,\phi,\theta^{\prime},\phi^{\prime})$\\
Groups of BSH coefficient & Three for R, G, B attributes & Two for real and imaginary parts\\
\bottomrule
\end{tabular}
\label{tab:compact_compare}
\end{table*}

Our BiWGS method requires training data consisting of Tx-Rx position pairs, corresponding wireless channel, and associated spatial spectrum. Notably, we use the spatial spectrum to facilitate the training. It characterizes the angular power distribution of the wireless channel.

Given the angular resolution $z$ and $v$ for azimuth and elevation angles, the spatial spectrum is defined as
\begin{equation}
    \boldsymbol{I}=\begin{bmatrix}
    P(\theta_1,\phi_1) & \cdots & P(\theta_1,\phi_z) \\
    \vdots & \ddots & \vdots \\
    P(\theta_v,\phi_1) & \cdots & P(\theta_v,\phi_z)
    \end{bmatrix},
\label{spectrum}
\end{equation}
where $P(\theta,\phi)$ denotes the power of the received signal at a certain AOA, which is generated via the Conventional Beamforming (CBF) (also called Bartlett beamforming) method \cite{bartlett1950periodogram} for its simplicity, which is expressed as
\begin{equation}
P(\theta,\phi) = \Vert \boldsymbol{b}^H(\theta,\phi)\boldsymbol{h} \Vert_2^2.
\label{spatial power}
\end{equation}

\begin{figure}[t]
    \centering
    \subfloat[linear]{
        \includegraphics[width=2.25cm,height=2.25cm]{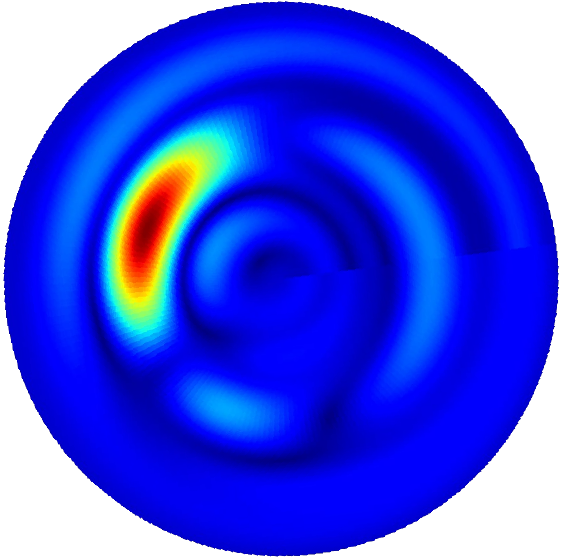}
        \label{linear_spectrum}
    }
    \hspace{0.1\linewidth} 
    \subfloat[dB-scale]{
        \includegraphics[width=2.25cm,height=2.25cm]{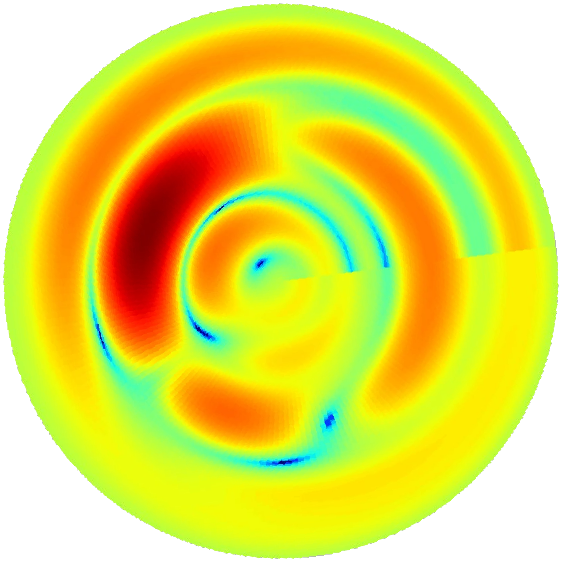}
        \label{fig:debris}
    }
    \caption{Comparison of linear and dB-scale spatial spectrum (a): linear spectrum. (b): dB-scale spectrum.}
    \label{spectrum_example}
\end{figure}

In the spatial spectrum, high-intensity regions (red in Fig.~\ref{linear_spectrum}) indicate stronger power at the corresponding AoA, suggesting a possible LOS path. However, the amplitude of LOS path surpasses that of scattering path by several orders of magnitude. Therefore, training on linear-scale spectrum introduces a bias toward the dominant LOS path, thereby suppressing contributions of weaker paths. This bias significantly degrades the inference performance of the proposed BiWGS model in NLOS scenarios. To mitigate this limitation, we apply a logarithmic transformation, converting the spectrum into the dB scale, which enhances the model’s ability to capture bidirectional scattering patterns. Fig.~\ref{spectrum_example} illustrates the difference between linear-scale and dB-scale spatial spectrum. It is notable that the spatial spectra in Fig.~\ref{spectrum_example} are the polar coordinates representation of spatial spectrum defined in (\ref{spectrum}) to facilitate visualization. In this representation, the radial coordinate is discretized into 
$v$ concentric rings, while the angular coordinate is sampled at $z$ points along each ring, corresponding to elevation and azimuth resolutions, respectively.

Furthermore, the loss function is designed as the mixture of spectrum loss $\mathcal{L}_s$ and channel power gain loss $\mathcal{L}_g$.
\begin{itemize}[leftmargin=*, noitemsep]
    \item \textbf{Spectrum loss:} Spectrum loss is defined by the $\mathcal{L}_2$ loss in term of mean square error (MSE) between the dB-scale ground-truth spectrum $\boldsymbol{I}_{\text{gt}}^{\text{dB}}$ and predicted spectrum $\boldsymbol{I}_{\text{pred}}^{\text{dB}}$, which is expressed as
    \begin{equation}
    \mathcal{L}_s = \mathcal{L}_2~(\boldsymbol{I}_{\text{gt}}^{\text{dB}},\boldsymbol{I}_{\text{pred}}^{\text{dB}}),
    \label{spectrum loss}  
    \end{equation}
    where $\boldsymbol{I}$ is determined by (\ref{spectrum}).

    \item \textbf{Channel power gain loss:} The dB-scale channel power gain is expressed as
    \begin{equation}
    g^{\text{dB}}_i = 10\log_{10}(\|\boldsymbol{h}_i\|_2^2), i\in\{\text{gt},\text{pred}\}.
    \label{channel}  
    \end{equation}
    Accordingly, channel power gain loss is defined by the $\mathcal{L}_1$ loss, which is the mean absolute error (MAE) between the ground-truth channel power gain $g_{\text{gt}}^{\text{dB}}$ and the predicted channel power gain $g_{\text{pred}}^{\text{dB}}$ as follows
    \begin{equation}
    \mathcal{L}_g = \mathcal{L}_1~(g_{\text{gt}}^{\text{dB}},g_{\text{pred}}^{\text{dB}}).
    \label{channel gain loss}  
    \end{equation}
\end{itemize}

Finally, the loss function is defined as the weighted sum of the above two losses, which is expressed as
\begin{equation}
\mathcal{L}= \eta_1\mathcal{L}_s + \eta_2\mathcal{L}_g,
\label{loss function}
\end{equation}
where $\eta_1$ and $\eta_2$ are the weighting coefficients to control the importance of different components of the loss function.

During the training process, the parameters of every Gaussian ellipsoid, including mean vector $\boldsymbol{\mu}$, rotation matirx $\boldsymbol{R}$, scaling matrix $\boldsymbol{S}$, maximum opacity $\alpha^{\text{max}}$, BSH coefficents $a^{\text{Re}}_{i,k}$ and $a^{\text{Im}}_{i,k}$ for $i,k\in \{1,\ldots,D+1\}$, maximum length of equivalent path $\gamma^{\text{max}}$, and angle-independent coefficient $Z$, are optimized via stochastic gradient descent using the adaptive moment estimation (Adam) optimizer. In addition, we apply the adaptive density control strategy in the training stage similarly as 3DGS illustrated in Section \ref{optical implementation detail} to control the numbers and size of Gaussian ellipsoids within the environment. Fig.~\ref{pipeline} illustrates the overall pipeline of the proposed BiWGS method. Furthermore, several major differences between the BiGS and BiWGS are summarized in Table \ref{tab:compact_compare}. 

\begin{table*}[t]
\centering
\caption{Comparison of 3D CKM Construction Performance}
\begin{tabular}{|c|c|c|c|c|c|c|c|c|}
\hline
\multirow{2}{*}{\textbf{Method}} & \multicolumn{2}{c|}{\textbf{Conference room}} & \multicolumn{2}{c|}{\textbf{Bedroom}} & \multicolumn{2}{c|}{\textbf{Office}} & \multicolumn{2}{c|}{\textbf{Average}} \\
\cline{2-9}
& \textbf{SSIM}$^{\uparrow}$ & \textbf{LPIPS}$^{\downarrow}$ & \textbf{SSIM}$^{\uparrow}$ & \textbf{LPIPS}$^{\downarrow}$ & \textbf{SSIM}$^{\uparrow}$ & \textbf{LPIPS}$^{\downarrow}$ & \textbf{SSIM}$^{\uparrow}$ & \textbf{LPIPS}$^{\downarrow}$ \\
\hline
NeRF\textsuperscript{2} \cite{zhao2023nerf2} 
& 0.5977 & 0.5718 & 0.7138 & 0.5477 & 0.6911 & 0.5186 & 0.6675 & 0.5460 \\
WRF-GS \cite{wen2025wrf} 
& 0.6919 & 0.5616 & 0.7210 & 0.5217 & 0.6444 & 0.6313 & 0.6858 & 0.5715 \\
\textbf{Proposed BiWGS} 
& 0.6610 & 0.4396 & 0.7002 & 0.4168 & 0.6748 & 0.5131 & 0.6787 & \textbf{0.4565} \\
\hline
\end{tabular}
\label{3D_table}
\end{table*}

\begin{figure*}[t!]
\centering
\includegraphics[width=0.9\linewidth]{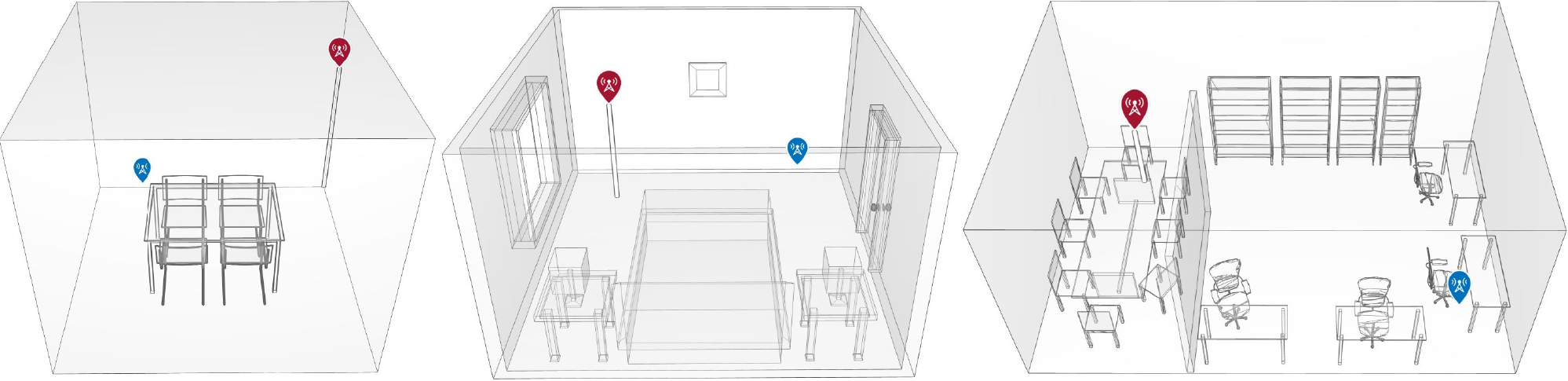}
\caption{Three 3D physical environments of datasets; conference room (left), bedroom (center), and office (right).}
\label{dataset}
\end{figure*}

\section{Experiment Results}
\label{results}
In this section, we evaluate the performance of our proposed BiWGS algorithm for 6D CKM construction. Specifically, we use three synthesis 3D scenes in \cite{lu2024newrf} as our 3D physical environment, shown in Fig.~\ref{dataset}. Moreover, we utilize the NVIDIA Sionna ray-tracing simulator \cite{hoydis2023sionna} to generate our datasets within the 3D physical environment. In our simulations, the frequency of signals is set as 6 GHz, with all transmission paths limited to a maximum scattering/reflection times of 3. The Rx is equipped with a half-wavelength UPA antenna configuration with $N_v = N_h = 4$. The azimuth and elevation resolutions of the spatial spectrum are set to $z=180$ and $v=180$, respectively, with resolution of 1$^\circ$. Lastly, taichi \cite{hu2019taichi} is used to implement parallelized computations in the CUDA kernels.

We consider both 3D and 6D CKM construction for performance comparison\footnote{BiWGS can be easily implemented for 3D CKM construction by maintaining either the Tx or Rx position fixed. As the SOTA approaches like NeRF$^2$ and WRF-GS are only applicable for 3D CKM construction, we consider BiWGS for 3D CKM construction as a use case to show the performance of  BiWGS in achieving high-fidelity CKM construction.}. In the 3D CKM construction experiments, all datasets comprise channel measurements acquired at a fixed Rx position with uniformly sampled Tx positions. Each dataset undergoes 90\%–10\% training-test partitioning. Moreover, the performance is evaluated by the quality of predicted spatial spectrums. Specifically, we use two evaluation metrics including structural similarity index measure (SSIM) and learned perceptual image patch similarity (LPIPS). Compared with SSIM, LPIPS incorporates spatial ambiguities to describe high-dimensional feature similarities between spectra \cite{zhang2018unreasonable}. In contrast, for the 6D CKM construction experiment, all datasets consist of two distinct sets of channel measurements: a training set containing measurements from 9 distinct Tx positions, and a test set containing measurements from a different Tx position (unseen in the training set). Rx positions at both the training set and the test set are uniformly sampled. Moreover, an 80\%–20\% training-to-test data ratio is maintained across all datasets. Performance for 6D CKM construction is evaluated based on the accuracy of the channel power gain prediction, quantified by the MAE and normalized mean absolute error (NMAE). Furthermore, all Tx and Rx maintain a minimum distance of one wavelength from 3D environmental objects to avoid reactive near-field region interactions\cite{liu2023near}. Any Tx or Rx violating this criterion will be discarded from the dataset. 

\begin{figure}[t]
\centering
\includegraphics[width=7.5cm,height = 7.5cm]{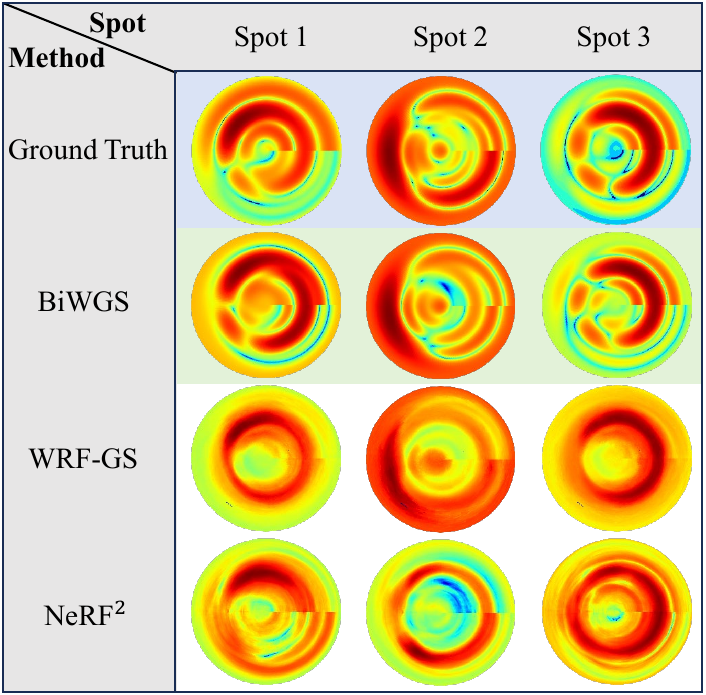}
\caption{Comparative visualization of spatial spectrum predictions for conference room.}
\label{Spectrum Comparison Conf}
\end{figure}

\begin{figure}[t]
\centering
\includegraphics[width=7.5cm,height = 7.5cm]{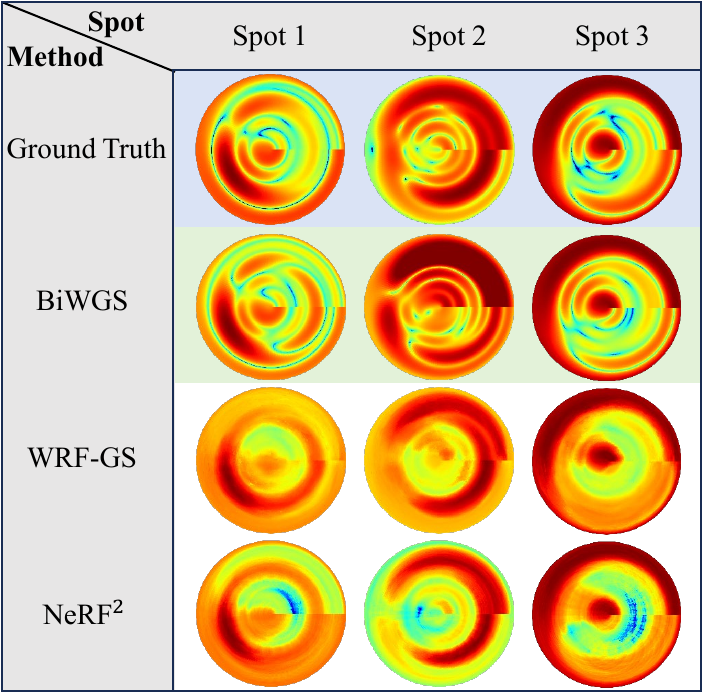}
\caption{Comparative visualization of spatial spectrum predictions for bedroom.}
\label{Spectrum Comparison Bedroom}
\end{figure}

\begin{figure}[t]
\centering
\includegraphics[width=7.5cm,height = 7.5cm]{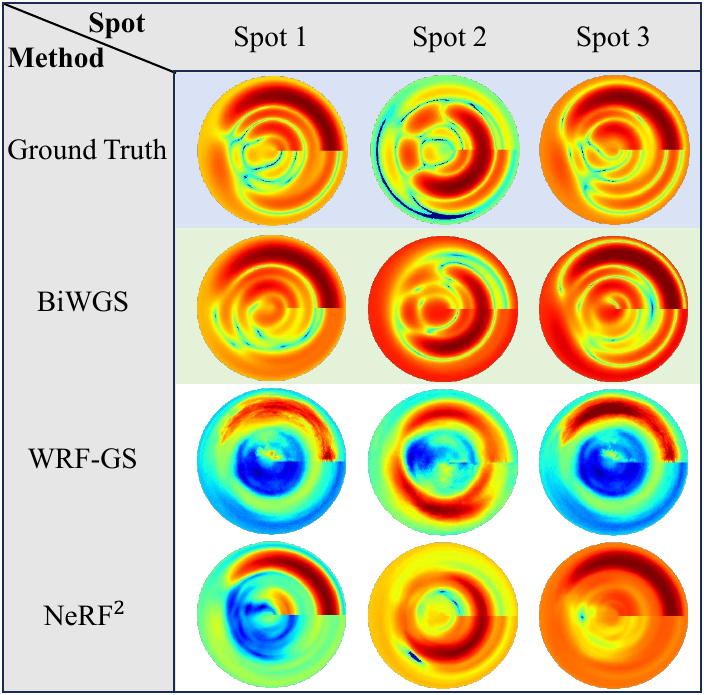}
\caption{Comparative visualization of spatial spectrum predictions for office.}
\label{Spectrum Comparison office}
\end{figure}

Table. \ref{3D_table} compares the median SSIM and LPIPS metrics of our method for 3D CKM construction, versus the benchmark schemes NeRF$^2$\cite{zhao2023nerf2} and WRF-GS \cite{wen2025wrf} at three different scenarios. On average, the median SSIM of BiWGS, WRF-GS, and NeRF$^2$ are 0.6787, 0.6858, and 0.6675, respectively. While the median LPIPS of BiWGS, WRG-GS, and NeRF$^2$ are 0.4565, 0.5715, and 0.5460, respectively. Consequently, the proposed BiWGS achieves a comparable performances as the SOTA method WRF-GS at the SSIM metric (with only a 0.007 gap). Moreover, the proposed BiWGS reaches the SOTA performance at the LPIPS metric. This is due to the fact that our explicit, bidirectional modelling can capture the features of the wireless transmission environment better than the implicit, unidirectional model WRF-GS or NeRF$^2$. Furthermore, Figs.~\ref{Spectrum Comparison Conf}, \ref{Spectrum Comparison Bedroom}, and \ref{Spectrum Comparison office} provide a visual comparison of the predicted spatial spectra among three methods and the ground truth across three scenarios. In each scenario, three example spots, corresponding to distinct Tx positions, are presented for illustration. The results show that both WRF-GS and NeRF$^2$ exhibit blurred spectral patterns with evident spatial ambiguities, whereas the proposed BiWGS method produces spectra with substantially fewer ambiguities. This improvement highlights the superior capability of BiWGS in preserving spectral fidelity, which is consistent with its advantage under the LPIPS metric.
\begin{table}[t]
\centering
\caption{Channel Power Gain Prediction Performance for 6D CKM Construction Across Environments}
\begin{tabular}{|c|c|c|c|}
\hline
Scenarios & Metric & MLPs\cite{saito2019two}  & \textbf{BiWGS} \\
\hline
\multirow{2}{*}{Conf. room} 
& MAE & 4.40 dB & \textbf{3.68 dB} \\
& NMAE & 0.096 & \textbf{0.080} \\
\hline
\multirow{2}{*}{Bedroom} 
& MAE & 7.81 dB & \textbf{4.93 dB} \\
& NMAE & 0.158 & \textbf{0.101} \\
\hline
\multirow{2}{*}{Office} 
& MAE & 14.60 dB & \textbf{6.70 dB}\\
& NMAE & 0.237 & \textbf{0.109} \\
\hline
\end{tabular}
\label{6D_table}
\end{table}

Table. \ref{6D_table} compares the 6D CKM construction performance between the proposed BiWGS method and the classical MLP-based approach\cite{saito2019two} for channel power gain prediction. Note that NeRF$^2$ and WRF-GS are not applicable for 6D CKM construction here. The results demonstrate that the proposed BiWGS approach exhibits a significant performance advantage in 6D CKM construction. Furthermore, the results indicate that BiWGS exhibits strong transferability, learning electromagnetic transmission characteristics from known Tx configurations and achieving high-accuracy channel power gain predictions at novel, unobserved Tx positions.

\section{Conclusion}
\label{conclusion}
This paper proposes BiWGS, a novel 6D CKM construction method inspired by the optical BiGS architecture. Our proposed method learns the bidirectional
scattering patterns of Gaussian ellipsoids to accurately fit the electromagnetic transmission characteristics of the wireless environment, thereby enabling the construction of 6D CKM. Comprehensive experiment evaluations demonstrate that BiWGS achieves spatial spectrum prediction accuracy comparable to SOTA 3D CKM construction techniques while also supporting 6D CKM construction. This represents a dimensionality expansion without compromising prediction fidelity. Some interesting directions for future extensions include computational complexity reduction, cross-frequency wideband 6D CKM construction, and BiWGS's applications.


%


\ifCLASSOPTIONcaptionsoff
  \newpage
\fi



%

\bibliographystyle{ieeetr}

\bibliography{IEEEabrv,ref}

\end{document}